\documentclass[11pt,a4paper]{article}
\pdfoutput=1
\usepackage{jheppub}
\usepackage[utf8]{inputenc}
\usepackage{graphicx,fancybox,float,comment}
\usepackage{units}
\usepackage{multirow,array,arydshln}
\usepackage{xcolor}
\usepackage{amsmath,amssymb}
\usepackage{cleveref}
\usepackage{yfonts}
\usepackage{tensor}
\usepackage{paralist}
\usepackage{braket}
\usepackage{tensor,mathrsfs}
\usepackage{ulem}

\title{Entropy production far from equilibrium in a chiral charged plasma in the presence of external electromagnetic fields}
\author{Casey Cartwright}

\affiliation{Department of Physics and Astronomy, University of Alabama,\\ University Blvd., Tuscaloosa, AL 35487, U.S.A. }

\emailAdd{cccartwright@crimson.ua.edu}

\begin{document}
\newcommand{\eref}[1]{Eq.~(\ref{#1})}
\newcommand{\exd}{\mathrm{d}}
\newcommand{\reals}{\mathbb{R}}
\newcommand{\mB}{\mathcal{B}}
\newcommand{\pd}{\partial}
\newcommand{\s}{\sigma}
\newcommand{\ads}{AdS$_5$ }
\newcommand{\mbf}[1]{\mathbf{#1}}
\newcommand{\mE}{\mathcal{E}}
\abstract{We report on the time evolution of a charged strongly coupled $N=4$ SYM plasma with an axial anomaly subjected to strong electromagnetic fields. The evolution of this plasma corresponds to a fully backreacted asymptotically AdS$_5$ solution to the Einstein-Maxwell-Chern-Simons theory. We explore the evolution of the axial current and production of axial charges. As an application we show that after a sufficiently long time both the entropy and the holographic entanglement entropy of a strip-like topology ( both parallel to and transverse to the flow of axial current) grow linearly in time.}

\maketitle

\section{Introduction}
It is expected that extremely large magnetic fields are generated during the collisions of heavy ions which produce a QGP~\cite{Kharzeev:2007jp,Skokov:2009qp}. \footnote{Standard lore states that the magnetic fields generated during collisions will not last long enough to produce an observable effect. } At high energy chiral symmetry is restored in the QCD Lagrangian leading to the presence of a chiral anomaly. This has led to the proposal of possible anomalous effects which might be seen on an event by event basis during the generation of a QGP, such as the chiral magnetic effect (CME)~\cite{Kharzeev:2004ey,Vilenkin1978}. The CME is due to the asymmetry between the number of particles and antiparticles with right handed and left handed helicity. And it can be shown that when one applies a magnetic field to such a system an electromagnetic current is generated in the direction of the magnetic field~\cite{Warringa:2012bq}. An observable two point correlation sensitive to the CME effect was first proposed in~\cite{Kharzeev:2004ey,Voloshin2004}. \footnote{Another observable was recently proposed in~\cite{Tang:2019pbl} (see also~\cite{Lin:2020jcp}).} By studying the azimuthally asymmetric distribution of charged hadron production both the STAR collaboration at RHIC and the ALICE collaboration at the LHC have observed the predicted fluctuation~\cite{Kharzeev2019,STAR2009CME,Deng:2016wt}. However the measurement may be obscured by the background with the geometry of the collision responsible for the observation. To correct for this, efforts are currently under way at RHIC with a dedicated isobar (nuclei with the same mass numbers and size but different electric charge) run~\cite{Skokov:2016yrj,Deng2016}. In condensed matter physics the effect has already been found to exist in Dirac semi-metals~\cite{Xiong413}. As this still remains inconclusive in heavy ion collisions it therefore motivates further study of thermalizing strongly coupled systems with a chiral anomaly. 

A powerful method for obtaining information about strongly coupled systems is via holography. There is a vast amount of literature\footnote{The references we cite here are only a small sample.} dedicated to thermalization~\cite{AbajoArrastia:2010yt,Balasubramanian:2011ur,Caceres:2012em,Ebrahim:2010ra,Keranen:2011xs,Camilo:2014npa,Hu:2016mym,Giordano:2014kya,Zhang:2015dia,Galante:2012pv,Dey:2015poa,Arefeva:2012jp,Atashi:2016fai,Zhang:2014cga,Ageev:2017wet,Andrade:2016rln,Wondrak:2017kgp,Wondrak:2020tzt}, often utilizing Vaidya spacetime, and to the study of dynamical holographic systems as analogues for heavy ion collisions~\cite{Chesler:2008hg,Chesler:2010bi,vanderSchee:2012qj,vanderSchee:2013pia,Casalderrey-Solana:2013aba,Casalderrey-Solana:2016xfq,Chesler:2015wra,Grozdanov:2016zjj,Waeber:2019nqd,Muller:2020ziz}. These studies simulate the evolution of SYM plasmas via numerical evolution of bulk Einstein equations and are meant to mimic the conditions of heavy ion collisions. The majority of these studies have been devoted to the collision of gravitational shock waves. However they have not included the time dependent magnetic fields which we know to be present during heavy ion collisions~\cite{Kharzeev:2007jp,Skokov:2009qp}. This is in part due to the difficulty of including even static magnetic fields in equilibrium. 

Early works with SYM plasma subjected to external magnetic fields were concerned with their thermodynamic properties~\cite{D'Hoker:2009mm,DHoker:2009ixq,DHoker:2010onp} with the first example of perturbative studies in~\cite{Janiszewski:2015ura}. Recently the importance of including magnetic fields in the study of SYM plasma for application to heavy ion collisions was demonstrated by showing the ratio of the transverse to longitudinal pressure ($P_T/P_L$) as a function of the $B/T^2$ agree between QCD and $N=4$ SYM plasma~\cite{Endrodi:2018ikq}. There have been two studies conducted in which the dynamical evolution of the Einstein equations include a fully back reacted magnetic field~\cite{Fuini:2015hba,Cartwright2019}. However both of these studies do not include a Chern-Simons term in the dual gravitational theory. This term, when included, provides for us an axial anomaly in the dual field theory. Other authors have utilized Vaidya spacetimes to include this term in a linearized analysis~\cite{Lin:2013sga,Pendas:2019}. In this work we make use of the techniques developed by~\cite{Chesler:2008hg,Chesler:2010bi,vanderSchee:2012qj,vanderSchee:2013pia,Casalderrey-Solana:2013aba,Casalderrey-Solana:2016xfq,Chesler:2015wra,Grozdanov:2016zjj,Waeber:2019nqd,Folkestad:2019lam,Muller:2020ziz,Fuini:2015hba,Cartwright2019} to extend the analysis of~\cite{Fuini:2015hba,Cartwright2019} to include the axial anomaly. This provides for us the simplest such setup in which to study the time-dependent relaxation of a far from equilibrium plasma with a chiral anomaly subjected to electromagnetic fields. It should be stressed that the electromagnetic fields created during a heavy ion collision are dynamically generated (i.\ e.\ a local gauge field). Our setup includes an external electric and magnetic field aligned along the $x_3$-direction (i.\ e.\ a global gauge field). In the presence of a chiral anomaly the aligned electric and magnetic field stimulates the production of axial charges. The increasing axial charge density contributes to the current density along the $x_3$-direction in which the chiral charges are accelerated by the electric field leading to Joule heating of the plasma. 

As an application of our numerical model we study the growth of entropy during the evolution. Entropy has been repeatedly shown to be a meaningful quantity to compare to experiments (some examples~\cite{Gubser:2008pc,Gubser:2009sx,Lin:2009pn}). An interesting aspect of the CME is that it produces a dissipation-less~\cite{Kharzeev:2011ds} current and hence does not contribute to thermal entropy production~\cite{Kharzeev:2007jp}.  Despite this lack of thermal or classical entropy production of the current associated with the CME we may expect there is a notion of entropy production due to the anomalous production of axial charges. 
Our results demonstrate that the production and subsequent acceleration of axial charges by the electric field produces a linear growth in the entropy. In addition we also compute the entanglement entropy in the dual field theory via methods used in~\cite{Ecker:2015kna,Ecker:2016thn,Cartwright2019}. We also find linear growth of the reduced entropy of strip like subsystems extending in directions both transverse to and parallel to the axial current flow.

Our work is divided as follows. We begin in~\cref{sec:Holo_Description} by introducing our the holographic description of our system. We then discuss the asymptotic analysis and introduce the dual energy-momentum tensor and current for our system. In~\cref{sec:Numerical_Tech} we briefly discuss the numerical techniques used to construct solutions to the Einstein equations. In~\cref{sec:Results} we display for the first time the energy-momentum tensor of a strongly coupled far from equilibrium charged plasma with chiral anomaly subjected to external electromagnetic fields. We also display for the first time the dynamical evolution of the axial current and axial charge density. Finally we investigate a simple application of our work by investigating the entropy production during the evolution. We compare the evolution of the thermal and entanglement entropy during the evolution with and without the production of axial charges.

\section{Setup}

\subsection{Holographic Description}\label{sec:Holo_Description}
We employ the characteristic formulation of general relativity first formulated in~\cite{Bondi:1960jsa,Sachs:1962wk} and implemented in a myriad of subsequent publications for the study of dynamical systems in asymptotically anti-de-Sitter spacetime (some examples~\cite{Chesler:2008hg,Chesler:2010bi,vanderSchee:2012qj,vanderSchee:2013pia,Casalderrey-Solana:2013aba,Casalderrey-Solana:2016xfq,Chesler:2015wra,Grozdanov:2016zjj,Waeber:2019nqd,Folkestad:2019lam,Muller:2020ziz,Cartwright2019}). The action for a bulk Einstein-Maxwell-Chern-Simons theory in five dimensions can be written as,
\begin{equation}
 S=-\int{\exd^5x\frac{1}{16\pi G_5}\left[ \sqrt{-g} (R-2 \Lambda 
-L^2F_{\mu\nu}F^{\mu\nu})\right]+\frac{k}{12\pi G_5}\epsilon^{\alpha\beta\gamma\delta\eta} 
\mathcal{A}_{\alpha}F_{\beta\gamma}F_{\delta\eta}},\label{CS-System}
\end{equation}
with $\epsilon^{\alpha\beta\gamma\delta\eta}$ the five dimensional totally antisymmetric Levi-Civita symbol ($\epsilon^{01234}=1$) and $\mathcal{A}$ is a $U(1)$ gauge field with $F=\exd\mathcal{A}$. The action also contains $G_5=\frac{\pi}{2}L^3/N_c^2$ the five dimensional Newton's constant, the cosmological constant $\Lambda$ which is related to the AdS radius $L$ via $\Lambda=-6/L^2$, the number of colors $N_c$ and the Chern-Simons coupling $k$ which will be written in a dimensionless form as $k=2\gamma\pi G_5$. The equations of motion which result from variation of the action are,
\begin{align}
R_{\mu\nu}-\frac{1}{2}g_{\mu\nu}R+\Lambda 
g_{\mu\nu}&=2L^2(F_{\mu\lambda}\tensor{F}{_{\nu}^{\lambda}}-g_{\mu\nu}\frac{1}{4}F_{\alpha\beta}F^{
\alpha\beta}), \\
\nabla_{\mu} F^{\mu\nu}&=\frac{k}{\sqrt{-g}L^2}\epsilon^{\nu\alpha\beta\lambda\sigma}F_{\alpha\beta}F_{\lambda\sigma}. \label{eqn:EOM}
\end{align}
We will work in units where \footnote{In~\cref{sec:scaling} we justify setting $L=1$ via a scaling relation. },
\begin{equation}
L=1,\qquad  \frac{1}{16\pi G_5}=1\label{eqn:units}
\end{equation}
for the remainder of this work to simplify the analysis.\\

\noindent\textbf{Einstein-Maxwell-Chern-Simons Equations:} To select an ansatz we consider the symmetries of our system. We wish to have aligned electric ($\vec{E}$) and magnetic ($\vec{B}$) fields in order to see the desired effect of the production of chiral charges. We choose to align both these fields along the $x_3$-direction. This breaks the $O(3)$-symmetry to an $O(2)$ in the $x_1-x_2$ plane. Additionally we expect the presence of a heat current along $x_3$, as a result we break the remaining parity symmetry in $x_3$ requiring a component of the metric $g_{t3}=g_{3t}\neq0$. With these symmetry considerations in mind the simplest ansatz for our desired setup is as follows,
\begin{equation}
 \exd s^2=\omega\exd v  +S(v,r)^2\left(e^{B(v,r)}\left(\exd x_1^2+\exd 
x_2^2\right)+e^{-2B(v,r)}\exd x_3^2\right), \label{eq:eddfinkmetric}
\end{equation}
with the one form $\omega=(-A(v,r)\exd v+F(v,r)\exd x_3+2\exd r)$. Our gauge field ansatz in radial gauge is of the form,
\begin{equation}
\mathcal{A}_{\mu}(v,r)=(0,\phi(v,r),\frac{1}{2}x_2\mathcal{B},-\frac{1}{2}x_1\mathcal{B},-
P(v,r)),\label{eq:gauge}
\end{equation}
with a constant magnetic field $\mathcal{B}$. 

Inserting our ansatz, eq.\ (\ref{eq:eddfinkmetric}) and eq.\ (\ref{eq:gauge}), into eq.\ (\ref{eqn:EOM}) the Maxwell equations reduce to three equations, the first two of which are displayed in eq.\ (\ref{eqn:Maxwell1}) and eq.\ (\ref{eqn:Maxwell2}) in terms of the bulk electric field $\mathcal{E}=-\partial_r\phi$,
\begin{align}
 0 &=\frac{e^{2 B(v,r)} S(v,r)^4 \left(F (v,r) \left(2 \partial_vB(v,r) \partial_rP(v,r)+\partial_v\partial_rP(v,r)\right)+\partial_v F(v,r) \partial_rP(v,r)\right)}{S(v,r)^3}\nonumber \\
 &+\frac{e^{2 B(v,r)} \partial_v S(v,r) S(v,r)^3 F (v,r) \partial_rP(v,r)+3 \partial_v S(v,r) S(v,r)^5 \mathcal{E} (v,r)}{S(v,r)^3} \nonumber \\
 &+\gamma  \mathcal{B}  \partial_v P(v,r)+\partial_v\mathcal{E}(v,r), \label{eqn:Maxwell1}\\
 0&=\frac{-e^{2 B(v,r)} S(v,r)^4 \left(F (v,r) \left(2 \partial_r B(v,r) \partial_rP(v,r)+\partial_r^2 P(v,r)\right)+\partial_r F(v,r) \partial_rP(v,r)\right)}{S(v,r)^3} \nonumber\\
  &+\frac{-e^{2 B(v,r)} \partial_r S(v,r) S(v,r)^3 F (v,r) \partial_rP(v,r)-3 \partial_r S(v,r) S(v,r)^5 \mathcal{E} (v,r)}{S(v,r)^3}\nonumber \\
 &-\gamma  \mathcal{B} \partial_rP(v,r)-\partial_r\mathcal{E}(v,r).\label{eqn:Maxwell2}
\end{align}
Despite their appearance these two equations have an analytic solution for the bulk electric field $\mathcal{E}$. The solution can be decomposed into a homogeneous ($\mathcal{E}_h$) and particular contribution ($\mathcal{E}_p$). The homogeneous contribution had been found by previous authors  in~\cite{Fuini:2015hba,Cartwright2019} when the Chern-Simons coupling is set to zero,
\begin{equation}
 \mathcal{E}_h(v,r)=\frac{\rho}{S(v,r)^3}.
\end{equation}
When the Chern-Simons term is not present the integration constant $\rho$ can be interpreted as the axial charge density (see eq.\ (\ref{eqn:Current})). This contribution to the total axial charge is a constant throughout the evolution and we are free to set this quantity to zero if we choose. The particular solution to eq.\ (\ref{eqn:Maxwell1}) and eq.\ (\ref{eqn:Maxwell2}) is given by,
\begin{equation}
 \mathcal{E}_p(v,r)=\frac{\mB\gamma P(v,r)+e^{2B(v,r)}S(v,r)F(v,r)\partial_rP(v,r)}{S(v,r)^3},\label{eq:maxwellsolHomo}
\end{equation}
and contains the dynamical contribution to the total axial charge as the system evolves. The total solution to eq.\ (\ref{eqn:Maxwell1}) and eq.\ (\ref{eqn:Maxwell2}) is the sum of the homogeneous and particular solutions,
 \begin{equation}
 -\partial_r\phi(v,r)=\mathcal{E}(v,r)=\frac{\rho+\mB\gamma P(v,r)+e^{2B(v,r)}S(v,r)F(v,r)P'(v,r)}{S(v,r)^3},\label{eq:maxwellsolParticular}
 \end{equation}
where a prime denotes differentiation with respect to the radial coordinate $r$. The solution given in eq.\ (\ref{eq:maxwellsolParticular}) can now be used in the Einstein equations which depend on $\mathcal{E}$.


The final Maxwell equation for $P$ cannot be solved without knowledge of the solutions for the metric components. We are required to include the equation for $P(v,r)$ in the set of Einstein equations to be solved numerically. Conveniently if one expresses the final Maxwell equation using the characteristic derivative the final Maxwell equation can be written as a first order ODE,
\begin{equation}
    \dot{P}'(v,r)=f_{\dot{P}}(\dot{P},\dot{B},\dot{S},F,S,P,B), \quad \dot{h}=\partial_th+\frac{1}{2}A\partial_rh,
\end{equation}
where here $f_{\dot{P}}$ is a source term which depends on the included metric components and their radial or dotted derivatives. Utilizing the characteristic derivative and including the final Maxwell equation into the characteristic Einstein equations we find the equations take the following form,
\begin{subequations}
\begin{align}
S''(v,r)&=f_S(S,P,B),\label{eqn:Characteristic_S} \\
F''(v,r)&= f_{F}(F,S,P,B),\label{eqn:Characteristic_chi}\\
\dot{S}'(v,r)&= f_{\dot{S}}(\dot{S},F,S,P,B),\label{eqn:Characteristic_sdot}\\
\dot{P}'(v,r)&= f_{\dot{P}}(\dot{P},\dot{B},\dot{S},F,S,P,B),\label{eqn:Characteristic_pdot}\\
\dot{B}'(v,r)&= f_{\dot{B}}(\dot{P},\dot{B},\dot{S},F,S,P,B),\label{eqn:Characteristic_Bdot} \\
A''(v,r)&= f_{A}(A,\dot{P},\dot{B},\dot{S},F,S,P,B), \label{eqn:Characteristic_A}\\
\dot{F}'(v,r)&= f_{\dot{F}}(\dot{F},A,\dot{P},\dot{B},\dot{S},F,S,P,B),\label{eqn:Characteristic_chidot} \\
\ddot{S}(v,r)&=f_{\ddot{S}}(\dot{F},A,\dot{P},\dot{B},\dot{S},F,S,P,B). \label{eqn:Characteristic_sdotdot}
\end{align}
\end{subequations}
The full equations are included in appendix~\ref{sec:Appendix_eqns}. Inspecting these equations one finds that eq.\ (\ref{eqn:Characteristic_S}) is no longer a linear ODE, the first equation of the nested list structure has developed a non-linearity by the inclusion of the Chern-Simons term and now requires two pieces of initial data, the anisotropy profile at the initial time $v_0$, $B(v_0,r)$, and the bulk electric field profile $P(v_0,r)$. In addition we find the equations for $\dot{P}$ and $\dot{B}$ fail to nest and must be solved simultaneously.\\

\noindent\textbf{Entanglement Entropy:} We employ Ryu-Takayanagi's conjecture for the entanglement entropy (EE)~\cite{Ryu:2006bv,Lewkowycz:2013nqa}. The entanglement entropy $S_A$ for a subsystem A of a CFT in $\reals^{3,1}$ is defined as,
\begin{equation}
 S_A=\frac{A(\gamma_A)}{4G_{5}}\label{eq:RT},
\end{equation}
where $A(\gamma_A)$ is the ``area'' of a 3 dimensional static minimal surface in AdS$_5$ with boundary $\pd A\subset\reals^{3,1}$. The area functional of the codimension 2 surface $\gamma_A$ in $AdS_5$ is,
\begin{equation}
  \mathscr{A}=\int{\exd^3\sigma\sqrt{\det\left(g_{\mu\nu}\frac{\pd\chi^{\mu}}{\pd \sigma^a}\frac{\pd\chi^{\nu}}{\pd\sigma^{b}}\right)}}\label{eq:area},
\end{equation}
where $\chi$ are the embedding coordinates of the surface and $\tilde{g}_{ab}=g_{\mu\nu}\frac{\pd\chi^{\mu}}{\pd \sigma^a}\frac{\pd\chi^{\nu}}{\pd\sigma^{b}}$ is the induced metric on this surface. Following the work of~\cite{Ecker:2015kna} we can specialize~\cref{eq:area} to the metric given in~\cref{eq:eddfinkmetric} for surfaces bounded by strips in the field theory aligned along the transverse $x_1(x_2)$ and longitudinal (or parallel) $x_3$ directions,
\begin{subequations}
\begin{align}
 S_{\perp}=\frac{\mathscr{A}_{\perp}}{V_{\perp}}&= \int{\exd\s \sqrt{-\dot{v}^2 A e^{-B} S^4-\frac{2 \dot{v} \dot{z} e^{-B} S^4}{z^2}+\dot{x_1}^2 S^6}},\\
  S_{\parallel}=  \frac{\mathscr{A}_{\parallel}}{V_{\parallel}}&= \int{\exd\s \sqrt{-A e^{2 B} S^4 \dot{v}^2+2 e^{2 B} F S^4 \dot{v} \dot{x_3}-\frac{2 e^{2 B} S^4 \dot{v} \dot{z}}{z^2}+S^6 \dot{x_3}^2}}.  
\end{align}\label{eq:EE}
\end{subequations}
Where $V_{\perp}=\frac{1}{ 4\pi}\int\exd x_2\exd x_3$ and $V_{\parallel}= \frac{1}{4\pi}\int\exd x_1\exd x_2$ are infinite volume contributions with which we measure with respect to. The expressions are essentially identical to those used in the case of colliding gravitational shock waves~\cite{Ecker:2016thn} where we have also suppressed the dependence of the metric components on the time and radial direction and represented $\exd Y(\s)/\exd\s=\dot{Y}$.

The areas we compute are divergent quantities which require regularization. To regulate our results for the time evolution we subtract the value for the entanglement entropy of empty AdS spacetime,
\begin{equation}
    \mathscr{A}-\mathscr{A}_{vacuum},
\end{equation}
as proven to be a valid regularization procedure in~\cite{Ecker:2016thn}.

\subsection{Asymptotic Analysis}
\label{sec:Asymptotic}
A near boundary solution is needed to extract field theory information. We seek solutions which asymptotically approach \ads as $r\rightarrow\infty$. This is the case if $g_{\mu\nu}(\mathbf{x},r)\rightarrow\eta_{\mu\nu}=\text{diag}(-1,1,1,1)$ as $r\rightarrow\infty$. Schematically this solution can be written as the following expansion provided we are in the appropriate coordinate system~\cite{Fuini:2015hba},
\begin{align}
    g_{\mu\nu}&\sim\eta_{\mu\nu}+\left(g^{(4)}_{\mu\nu}(\mathbf{x})+h^{(4)}_{\mu\nu}(\mathbf{x})\log(r/L)\right)\left(\frac{L^2}{r}\right)^4+\cdots \\
    \mathcal{A}_{\mu}&\sim \mathcal{A}^{(0)}_{\mu}(\mathbf{x})+\mathcal{A}^{(2)}_{\mu}(\mathbf{x})\left(\frac{L^2}{r}\right)^2+\cdots \label{eqn:Expand_A}
\end{align}
where $g_{\mu\nu}$ bulk spacetime metric and $\mathcal{A}_{\mu}$ is the bulk gauge field. Expanding the metric components in a power series around $r\rightarrow\infty$ we simultaneously solve the Einstein and Maxwell equations order by order and arrive at the asymptotic solution,
\begin{align}
 B(v,r)&= \frac{b_4(v)}{r^4}+\frac{-24 b_4(v) \xi(v)+6 v b_4'(v)+2 E^2 \xi(v)+2 \mathcal{B}^2 \xi(v)}{6 r^5}\nonumber \\
 &+\log(r)  \left(-\frac{5 \left(8 E^2 \xi(v)^3+8 \mathcal{B}^2 \xi(v)^3\right)}{6 r^7}+\frac{5 \left(4 E^2 \xi(v)^2+4 \mathcal{B}^2 \xi(v)^2\right)}{6 r^6}\right. \nonumber \\
 &\left.-\frac{2 \left(2 E^2 \xi(v)+2 \mathcal{B}^2 \xi(v)\right)}{3 r^5}+\frac{E^2+\mathcal{B}^2}{3 r^4}\right)+\cdots\label{eqn:nbeB},\\
S(v,r)&=r+\xi(v)-\frac{E^2}{18 r^3}+\frac{2 E^2 \xi(v)-8 E p_2(v)}{60 r^4}\nonumber\\
&+\frac{ \log(r) \left(-168 E^2 b_4( v)-168 \mathcal{B}^2 b_4(v)-44 E^2 \mathcal{B}^2-43 E^4-\mathcal{B}^4\right)}{1764 r^7}+\cdots\label{eqn:nbeS},\\
A(v,r)&=r^2+2 r \xi(v)-2  \xi '(v)+\xi(v)^2+\frac{a_4(v)}{r^2}\nonumber \\
&+ \log(r) \left(\frac{8 E^2 \xi(v)^3+8 \mathcal{B}^2 \xi(v)^3}{3 r^5}+\frac{-4 E^2 \xi(v)^2-4 \mathcal{B}^2 \xi(v)^2}{2 r^4}\right. \nonumber\\
&\left.+\frac{2 \left(2 E^2 \xi(v)+2 \mathcal{B}^2 \xi(v)\right)}{3 r^3}-\frac{2 \left(E^2+\mathcal{B}^2\right)}{3 r^2}\right)+\cdots \label{eqn:nbeA},\\
F(v,r)&=\frac{f_4(v)}{r^2}-f_4(v)\frac{\log(r ) \left(\mathcal{B}^2 +E^2\right)}{3 r^6} +\cdots\label{eqn:nbeF},\\
P(v,r)&=p_0+E v+\frac{E}{r}+\frac{p_2(v)}{r^2}+\frac{2 \left(E \mathcal{B}^2+E^3\right) \log(r)}{15 r^5}+\cdots\label{eqn:nbeP} ,\\
\phi(v,r)&=\mu(v)+\frac{(p_0\mB\gamma+E v\mB\gamma +\rho)}{2r^2}+\cdots, \label{eqn:nbephi}
\end{align}
where $\xi(v)$ is a residual diffeomorphism symmetry which is fixed during the computation. In all the above expansions the ellipses include higher order terms in $1/r$ and additional $\log(r)$ terms including powers of $\log(r)$. The coefficients $b_4(v), a_4(v), f_4(v), p_2(v),\mu(v)$ typically cannot be determined by a near boundary solution to the Einstein equations\footnote{In the present work we will show that conservation of the dual energy-momentum tensor (hydrodynamic equations of motion) will determine $f_4$.}. These coefficients can only be determined by a full solution to the system. The coefficients $b_4(v)$, $a_4(v)$ and  $f_4(v)$ will appear in the dual energy momentum tensor displayed in eq.\ (\ref{eqn:enmomE})-(\ref{eqn:enmomPl}) and are the holographic dual of the pressure anisotropy, the energy density and the heat current. The coefficient $p_2(v)$ enters in the dual $U(1)$ current displayed in eq.\ (\ref{eqn:Current}) and the coefficient $\mu(v)$ can be interpreted as the dual chemical potential and is calculated as displayed in eq.\ (\ref{eqn:ChemicalPotential}). In near boundary expansion displayed in eq.\ (\ref{eqn:nbeB})-(\ref{eqn:nbephi}) we have already conveniently chosen to name one of the coefficients $E$. We chose to use this symbol due to the identification of $E$ as the electric field in the dual field theory. We can see the appearance of an electric field in the dual field theory by investigating the zeroth order coefficient of $\mathcal{A}_{\mu}$ as given in eq.\ (\ref{eqn:Expand_A}). Using the near boundary expansion given in eq.\ (\ref{eqn:nbeB})-(\ref{eqn:nbephi}) the zeroth order coefficient of $\mathcal{A}_{\mu}$ is given as\footnote{The dual field theory coordinates are $(t,x_1,x_2,x_3)$.},
\begin{equation}
\mathcal{A}_{\mu}^{(0)}=(\mu(t),x_2\mathcal{B}/2,-x_1\mathcal{B}/2,-p_0-Et), \label{eqn:fieldtheoryGaugeField}
\end{equation}
and can be identified as the external (global $U(1)$ invariant) gauge field in the dual field theory. Computing the field strength $F_{\mu\nu}^{(0)}$ associated with $\mathcal{A}_{\mu}^{(0)}$ we find,
\begin{equation}
 F^{(0)}_{12}=-F^{(0)}_{21}=\mathcal{B},\hspace{2cm} F^{(0)}_{03}=-F^{(0)}_{30}=-E,
\end{equation}
where we now see in the field theory we have aligned electric and magnetic fields along the $x_3$ axis. The electric field has entered as the time dependent source, or $O(r^0)$ term, in the near boundary expansion of $P(v,r)$. The coefficient $p_0$ can be interpreted as a constant or reference momentum per unit charge~\cite{Konopinski:1978}. Our choice of $P(v,r)\sim p_0+E t+O(r)$ is analogous to choosing a linearly increasing chemical potential $\mu(t)=\mu_0+\mu_1 t$. The coefficient $\mu_0$ is a constant background chemical potential or energy per unit charge. Altogether our field theory gauge field/potential is describing a time dependent energy per unit charge (time component), a constant momentum per unit charge in both the $x_1$ and $x_2$ directions and a linearly increasing momentum per unit charge in the $x_3$ direction.  

Solving the Einstein equations near the conformal boundary also yields first order ODE's for the asymptotic coefficients for $f_4$ and $a_4$,
\begin{equation}
    f_4'(t)=E((p_0+E t)\mB\gamma+\rho)\qquad a_4'(t)=\frac{8E}{3}(\xi(t)E+p_2(t)).\label{eqn:energy}
\end{equation}
In our numerical scheme we provide initial values for the coefficients $f_4$ and $a_4$ and use eq.\ (\ref{eqn:energy}) to evolve forward in time. The ODE for $f_4$ in eq.\ (\ref{eqn:energy}) has an analytic solution given by,
\begin{equation}
    f_4(t)=E((p_0+\frac{E}{2} t)\mB\gamma+\rho)t+f^{(0)}_{4}.\label{eqn:patho}
\end{equation}
The solution for $f_4(t)$ in eq.\ (\ref{eqn:patho}) reveals our choice of including only an axial gauge field is partially pathological. The solution given in eq.\ (\ref{eqn:patho}) for the coefficient $f_4(t)$ will grow without bound. There are two reasons for the unbounded growth of the coefficient $f_4$ in our work. First our system is translationally invariant in the spatial directions. Without any inhomogeneity our plasma is essentially a perfect conductor~\cite{Blake:2013owa}. By breaking the translational symmetry one can introduce momentum relaxation and hence resistivity to the plasma, examples of the introduction of resistivity include massive gravity~\cite{Vegh:2013sk,Blake:2013owa} and Q-Lattice models~\cite{Donos:2014cya}\footnote{The author thanks the referee for pointing the author to the references~\cite{Blake:2013owa,Donos:2014cya}.}. The second reason for the unbounded growth of the coefficient $f_4(t)$ is the chiral anomaly. The anomaly will continuously produce axial charges at a rate proportional to $\vec{E}\cdot\vec{\mathcal{B}}$. Hence the anomalous production of axial charges in a system without momentum relaxation leads to unbounded growth of both the current and heat current. We can see the unbounded charge accumulation by calculating the total charge in the system. The expectation value of the time component of the current operator, $\braket{J^0}$, dual to the time component of the bulk gauge field, $\mathcal{A}_{0}$, is encoded in the coefficient $\mathcal{ A}^{(2)}_0$,
\begin{equation}
   \mathcal{ A}^{(2)}_0=2\left(\rho+\mB\gamma(p_0+E t)\right).\label{eqn:totalCharge}
\end{equation}
Indeed \cref{eqn:totalCharge} shows our homogeneous system with an axial anomaly leads to a total charge that grows without bound\footnote{ The definition of the charge density is a little more subtle in our current choice of Eddington-Finkelstein coordinates, the total charge density with our choice of units in eq.\ (\ref{eqn:units}) is displayed in eq.\ (\ref{eqn:Current}).}.  

Unlike the ODE for the coefficient $f_4(t)$, the ODE for $a_4(t)$ cannot be solved without knowledge of the full solution to the Einstein equations. The ODE for $a_4$ in \cref{eqn:energy} has two contributions. The first contribution, $a_4'\propto E\, \xi$, arises from the location of the apparent horizon changing the effective energy of the system. This contribution can be removed by working in a fixed frame $\xi=0$. The second contribution, $a_4'\propto E\, p_2$, is a Joule heating term. One can show that the non-equilibrium contribution to the dual current is $\braket{J_3}\propto p_2(t)$ (see eq.\ (\ref{eqn:Current})) and hence $a_4'\propto \braket{\vec{J}} \cdot \vec{E}$.

The field theory energy-momentum tensor can be computed by including the proper counter terms to the action and utilizing the near boundary expansion eq.\ (\ref{eqn:nbeB}) to eq.\ (\ref{eqn:nbephi}). We follow the same conventions set in~\cite{Fuini:2015hba} for the procedure of holographic renormalization (see~\cite{Skenderis:2008dg,Taylor:2000xw,DHoker:2009ixq,Fuini:2015hba}). 
In our choice of units, eq.\ (\ref{eqn:units}), this procedure yields the following boundary stress-energy tensor \footnote{The renormalization point dependence of the energy-momentum tensor was carefully discussed in~\cite{Fuini:2015hba} Displaying the results of our calculation requires a choice of $\mu_r$. A rather un-physical choice is $\mu_r=1/L$. A more detailed discussion of this choice in this system will be carried out in future work. Please also see~\cite{Grozdanov:2017kyl} for further discussion of these points in the context of generalized global symmetries in holography.}.
\begin{align}
 \braket{T_{00}}&=-3 a_4(t)-\frac{4 E^2}{3} +2( E^2 +\mB^2)\log (\mu_r  )\label{eqn:enmomE},\\
\braket{T_{03}}&=\braket{T_{30}}= 4f_4(t) \label{eqn:enmomCur},\\
\braket{T_{11}}&=\braket{T_{22}}=-a_4(t)+4b_4(t)-\frac{E^2}{9}-\mathcal{B}^2+2( E^2 +\mB^2)\log (\mu_r  ) \label{eqn:enmomPt},\\
\braket{T_{33}}&= -a_4(t)-8 b_4(t)+\frac{8 E^2}{9}-2( E^2 +\mB^2)\log (\mu_r  ). \label{eqn:enmomPl}
\end{align}
Computing the trace of this energy-momentum tensor gives the expected conformal anomaly,
\begin{equation}
\braket{\tensor{T}{^{\mu}_{\mu}}}=-F_{\mu\nu}F^{\mu\nu}=2\left(E^2-\mB^2\right).
\end{equation}
Following~\cite{DHoker:2009ixq} we can also extract the following global current using,
\begin{equation}
 - 4\pi G_5  \braket{J^{\mu}}=\lim_{r\rightarrow \infty}-r^3L^2\eta^{\mu\nu}\partial_rA_{\nu}+\frac{k}{3}\epsilon^{\mu\nu\alpha\beta}A_{\nu}F_{\alpha\beta}.
\end{equation}
Given the choice made in eq.\ (\ref{eqn:units}) the one point function of the axial current density is given by\footnote{This is the so-called consistent current, it contains the Bardeen-Zumino term.},
\begin{equation}
     \braket{J^{\mu}}=\lim_{r\rightarrow \infty}4r^3\eta^{\mu\nu}\partial_rA_{\nu}-\frac{\gamma}{6}\epsilon^{\mu\nu\alpha\beta}A_{\nu}F_{\alpha\beta}.\label{eqn:extract_current}
\end{equation}
Applying equation eq.\ (\ref{eqn:extract_current}) we find the following form of the dual current one point function,
 \begin{equation}
     \braket{J^{\mu}}=\left(\frac{11\left( p_0+Et\right) \mB \gamma}{3}+4\rho,-\frac{1}{6}E\mB\gamma x_1,-\frac{1}{6}E\mB\gamma x_2,8 p_2(t)-\frac{1}{3}\mB\gamma \mu(t)\right).\label{eqn:Current}
 \end{equation}
As expected the external electric field $E$ contributes to the total energy density of the field theory (see eq.\ (\ref{eqn:enmomE})). Along with the anomalous current flow there is a time dependent heat current $f_4(t)$ (see eq.\ (\ref{eqn:enmomCur})). The system is anisotropic with a transverse and longitudinal pressure (see eq.(\ref{eqn:enmomPt})-(\ref{eqn:enmomPl})). The external electric field in the $x_3$-direction provides a contribution to the pressure in the $x_1-x_2$ plane and $x_3$-direction. The source of this pressure contribution can be attributed again to the presence of the Chern-Simons coupling. The $x_3$ component of the current also contains both an equilibrium ($\frac{1}{3}\mB\gamma\mu(t)$ see~\cite{Ammon:2017ded}) and a non-equilibrium ($8p_2(t)$) contribution (see eq.\ (\ref{eqn:Current})). Where the chemical potential $\mu(t)$ can be calculated as~\cite{Fuini:2015hba},
\begin{equation}
 \mu(t)=\int_{r_h}^{\infty}\exd r \mathcal{E}(r,t). \label{eqn:ChemicalPotential}
\end{equation}
Finally both the $x_1$ and $x_2$ component of the current eq.\ (\ref{eqn:Current}) are non-zero. These components indicate there is a azimuthally symmetric inflow of axial charges. A further pathology can be seen in eq.\ (\ref{eqn:Current}), the current is proportional to unbounded coordinates $x_1$ and $x_2$. This contribution to the current will grow infinite at infinite distance. We might expect that this is in part due to our system being infinite in extent and a more reasonable calculation intended to model the evolution of a plasma in a ``box'' would alleviate these seemingly infinite contributions.

\section{Numerical Techniques}\label{sec:Numerical_Tech}
The numerical solution to the characteristic Einstein equations have been carefully described in many works~\cite{Chesler:2008hg,Chesler:2009cy,Chesler:2010bi,Chesler:2013lia,wilkethesis,Janik:2017ykj,Casalderrey-Solana:2013aba,Casalderrey-Solana:2016xfq,Cartwright2019,Waeber:2019nqd,Folkestad:2019lam} etc. for particularly nice treatments see~\cite{Chesler:2013lia,Waeber:2019nqd}. In addition the techniques used to compute the entanglement entropy have also 
been described in detail in~\cite{Ecker:2015kna,Ecker:2018jgh,Cartwright2019} with a particularly nice treatment in~\cite{Ecker:2018jgh}. With this in mind we will not describe in depth the methods of construction for these solutions. We will only give a brief statement of the methods used. 

Each of our radial differential equations is solved by means of a Chebyshev spectral method (for an introduction see~\cite{boyd}). In order to tame CFL instabilities we employed domain decomposition in the radial grid, typically using 6 sub-domains each with $N=24$ grid points (see~\cite{Waeber:2019nqd} for a quick explanation). The number of needed grid points is larger then that found in~\cite{Waeber:2019nqd} for instance. This is due to the presence of the logarithmic terms which appear due to the electric and magnetic field. These terms ruin the typical ``exponential'' convergence of a spectral scheme. 

In order to step forward in time we employed a standard 4th order Runga-Kutta scheme with a time step of the order $dt\approx \frac{1}{4N^2}$. Our system contains a thermalizing black brane so we use the residual diffeomorphism symmetry to fix the location of the apparent horizon during the evolution of our system. In our previous work~\cite{Cartwright2019} we followed a method provided in~\cite{wilkethesis}, calculating an explicit differential equation for $\xi$. However in this work we have changed this to something similar to what is done in~\cite{Waeber:2019nqd}, fixing the behavior of the metric function $A$ on the apparent horizon and extracting $\partial_t\xi$ from the near boundary behavior of $A$ via $\xi'(t)=\frac{-1}{2}A_s(t,z)|_{z=0}$. 

We will outline our solution algorithm since it differs slightly from previous works. In order to construct solutions we do the following.
\begin{enumerate}
    \item Fix $B(v,r)$ on the initial time step.
    \item Solve the linear equation given by eq.\ (\ref{eqn:Characteristic_S}) in the limit of vanishing Chern-Simons coupling and vanishing bulk electric field $P(v,r)$, for $S_{\text{Linear}}$.
    \item  Fix $P(v,r)$ on the initial time step and solve the nonlinear system for $S(v,r)$ using Frechet differentiation and Newton iteration. The linear solution $S_{\text{Linear}}$ serves as an initial guess.
    \item Solve eq.\ (\ref{eqn:Characteristic_chi}) and eq.\ (\ref{eqn:Characteristic_sdot}) in turn as a nested system.
    \item Solve eq.\ (\ref{eqn:Characteristic_pdot}) and eq.\ (\ref{eqn:Characteristic_Bdot}) as a coupled system.
    \item Solve eq.\ (\ref{eqn:Characteristic_A}) and eq.\ (\ref{eqn:Characteristic_chidot}) in turn as a nested system.
    \item Extract time derivatives $\partial_vB(v,r)$ and $\partial_vP(v,r)$ from the definition $\dot{h}=\partial_vh+\frac{1}{2}A(v,r)\partial_rh(v,r)$. 
    \item On the next time step use the previous solution to eq.\ (\ref{eqn:Characteristic_S}) as an initial guess for newton iteration of the non-linear system.
    \item Repeat steps 4-8 for the duration of the evolution.
\end{enumerate}
In order to begin our time evolution on the initial time step we must repeatedly follow steps 1-4 in order to fix the location of apparent horizon to a numerically convenient location. In our case we fix this location to be at $z_h=1$.

On each time step we choose to solve for ``subtracted'' functions $f_s$ rather then the full function $f$ by using the known behavior of the function near the AdS boundary (see section~\ref{sec:Asymptotic}). When the logarithmic terms are present in the near boundary solution this is a necessary step in order to achieve a stable evolution. As an example, for the function $B$ we write,
\begin{align}
 B(v,z)&=z^4 B_s(v,z)+\log \left(\frac{1}{z}\right) \left(-\frac{1}{3} 20 z^7\xi (v)^3 \left(E^2 +\mathcal{B}^2 \right) \right. \nonumber\\
 &\left. +\frac{10}{3} z^6 \xi (v)^2\left(E^2 +\mathcal{B}^2 \right)-\frac{4}{3} z^5 \xi (v)\left(E^2 +\mathcal{B}^2 \right)+\frac{1}{3} z^4 \left(E^2+\mathcal{B}^2\right)\right).\label{eqn:Bsub}
\end{align}
We do this for all functions, writing them as $f(v,z)=z^{\delta}f_s(v,z)+\Delta_f(v,z)$ and substitute the resulting expressions into the Einstein equations before evolution. The use of subtracted functions also provides a simple method of extracting the information needed to construct the one point functions. Considering again the function $B(v,z)$ we need the coefficient $b_4(t)$ to construct the energy-momentum tensor, this enters as the coefficient proportional to $z^4$ in the near boundary expansion, eq.\ (\ref{eqn:nbeB}). With the choice made in eq.\ (\ref{eqn:Bsub}) the coefficient $b_4$ can be computed simply as,
\begin{equation}
 b_4(t)= \lim_{z\rightarrow 0} B_s(v,z).
\end{equation}
The coefficient $p_2(t)$ can be extracted in a similar manner from $P_s(v,z)$.  Appendix~\ref{sec:Appendix_symm} provides further discussion of extracting the necessary near boundary coefficient of $P(v,r)$.

We utilize a relaxation method to compute solutions to the geodesic equation as done in~\cite{Ecker:2015kna,Cartwright2019} (a basic introduction can be found in~\cite{numericalrecipes}). We typically use 350 grid points to approximate the solutions. The method computes the geodesics on a cutoff surface located at $z_{UV}=.075$. The method takes empty conformal AdS geodesics as an initial guess on the first time step. Once a solution is found it serves as the guess on the next time step. We have verified that on the range $z_{UV}\in [0.05,0.1]$ for strip widths of $\ell=0.8$ the calculated value of the entanglement entropy is cutoff dependent on the order of $10^{-2}$ which is more then sufficient for our purposes.

\section{Results}\label{sec:Results}
\begin{figure}[!htb]
\minipage{0.48\textwidth}
  \includegraphics[width=7.25cm]{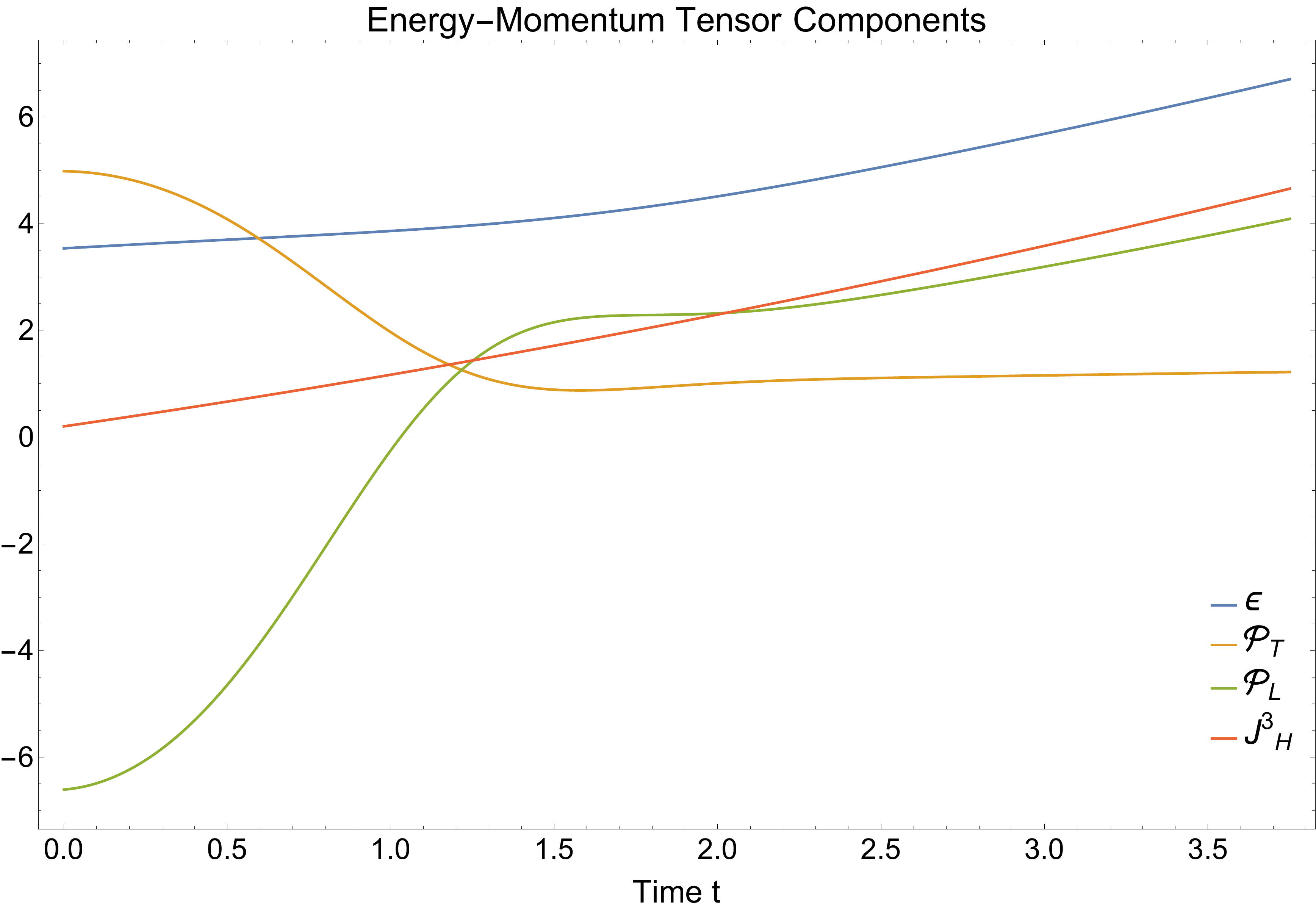}
\endminipage\hfill
\minipage{0.48\textwidth}
  \includegraphics[width=7.25cm]{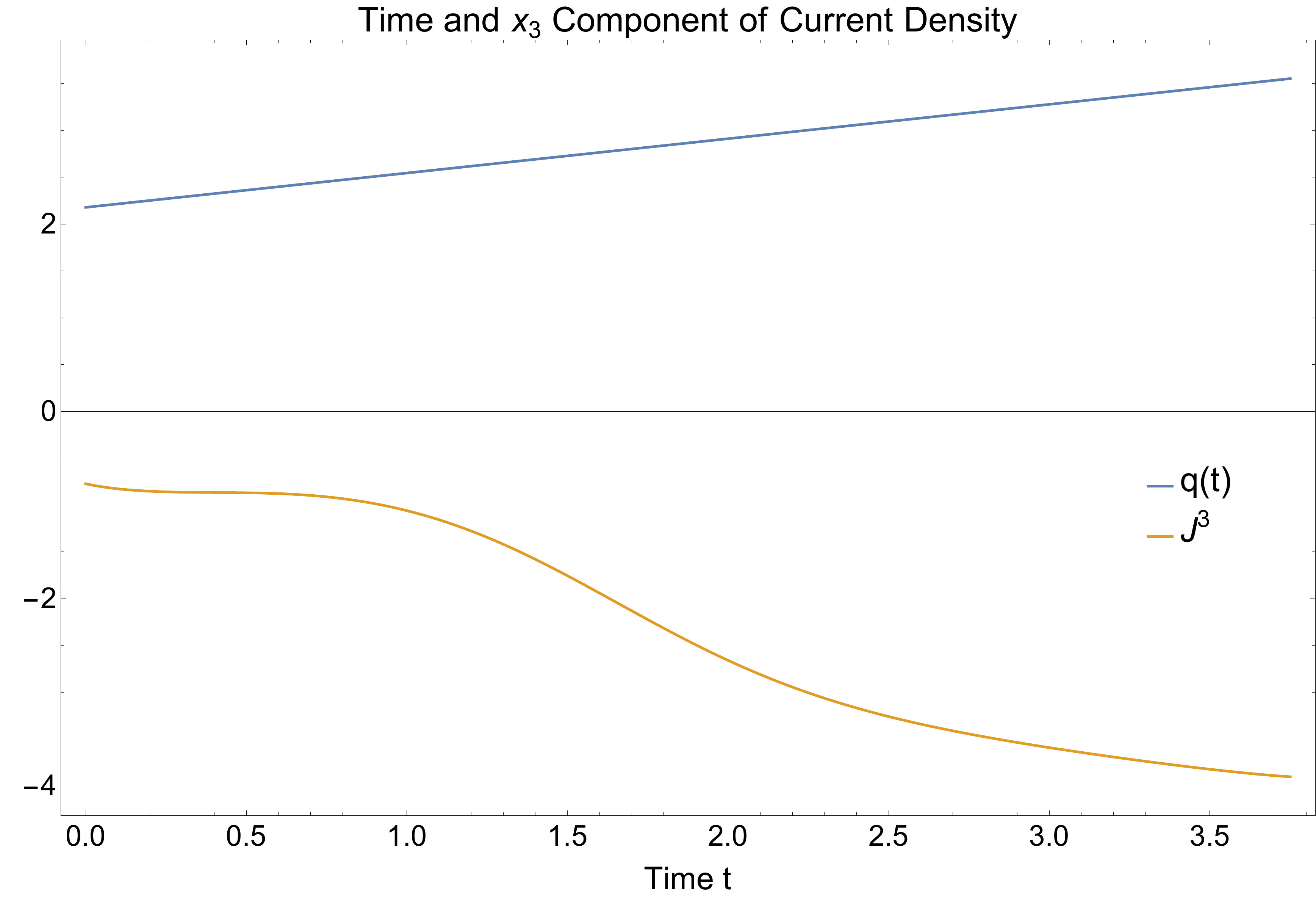}
\endminipage\hfill
  \caption{Time evolution of a strongly coupled far from equilibrium plasma with an axial anomaly subjected to an aligned external electric and magnetic field. \textit{Left:} The time evolution of the one point functions of the energy-momentum tensor are displayed at $\mu_r=1.2$. The blue line is the energy density $\epsilon=\braket{T_{00}}$, the green line the longitudinal pressure $\mathscr{P}_L=\braket{T_{33}}$, the orange line the transverse pressure $\mathscr{P}_T=\frac{1}{2}(\braket{T_{11}}+\braket{T_{22}})$ and the red line is the heat current $J^3_H=\braket{T_{03}}$. \textit{Right:} The time evolution of the time dependent components of the current density $\braket{J^{\mu}}$ are displayed. The blue line is the axial charge density $\braket{J^{0}(t)}$ and the orange line is the $x_3$ component of the current,  $\braket{J^{3}(t)}$.\label{fig:Enmom}}
\end{figure}

\textbf{Isotropization: }In figure~\ref{fig:Enmom} we display the non-zero components of the energy-momentum tensor along side the non-zero components of the axial current. For this evolution we chose to use the following form of the subtracted functions $B_s$ and $P_s$ at the initial time step $t=0$,\footnote{It should be noted that $v$ and $t$ coincide at the boundary $z=0$.}
\begin{equation}
    B_s(0,z)=e^{-z^2}, \quad P_s(0,z)=-\beta e^{-z^2},\label{eqn:initProfile}
\end{equation}
displayed here in the $\xi=0$ frame with $\beta=1/10$. In figure~\ref{fig:Enmom} we fix $\mB=1/2$, $\gamma=1/2$, $\rho=0.429$, $p_0=1/2$ and $E=2/5$. We begin the evolution with $a_4(v=0)=-5/4$ and $f_4(v=0)=5/100$. Although the initial time evolution is sensitive to the choice of the subtracted functions and the choice of parameters the general behavior of the late time evolution is not (see for example~\cite{wilkethesis}). More complicated initial radial profiles then eq.\ (\ref{eqn:initProfile}) can be considered and the resulting initial time evolution can be highly non-trivial. However when making a more complicated choice of initial data one must then separate the non-trivial initial time dynamics from other effects present during the evolution. In this work we choose not to do this, we consider the initial choice of data described above in order to clearly capture the essential physics.

In figure~\ref{fig:Enmom} we see the energy $\braket{T_{00}}$ of the solution continues to grow as an increasing number of axial charges are produced by the anomaly and subsequently accelerated by the electric field. As the total number of charges grows so does the $x_3$ component of the dual current $\braket{J_{3}}=8p_2(t)-\mB\gamma\mu(t)/3$ and the heat current  at the boundary $\braket{T_{03}}=4f_4(t)$. The transverse $(\braket{T_{11}}+\braket{T_{22}})/2$ and longitudinal pressures $\braket{T_{33}}$  oscillate as they undergo the isotropization process. However the continuous growth of the energy can be seen overtaking the isotropization process. It is interesting to note that while the energy density is increasing the transverse pressure at late times stays roughly constant. It is the longitudinal pressure which grows in order to satisfy the trace condition on the energy-momentum tensor. This may have been expected considering the continued growth of the $x_3$ component of the current density. Our work can be compared to previous work~\cite{Fuini:2015hba,Cartwright2019} which demonstrates that without the continuous production of axial charges the transverse and longitudinal pressures relax to a final anisotropic state due to the presence of the magnetic field. 

\begin{figure}[!htb]
\minipage{0.32\textwidth}
  \includegraphics[width=4.75cm]{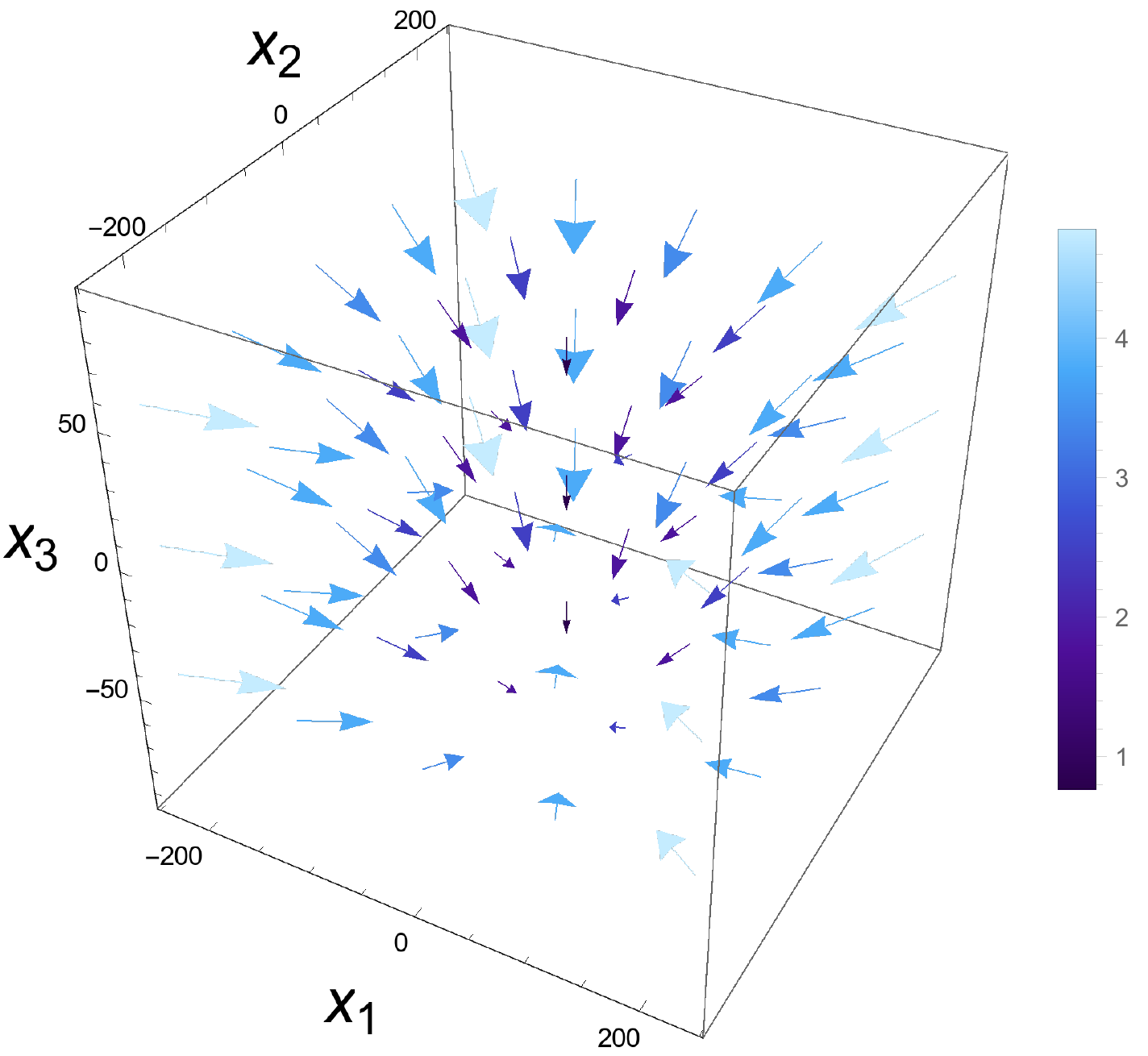}
\endminipage\hfill
\minipage{0.32\textwidth}
  \includegraphics[width=4.75cm]{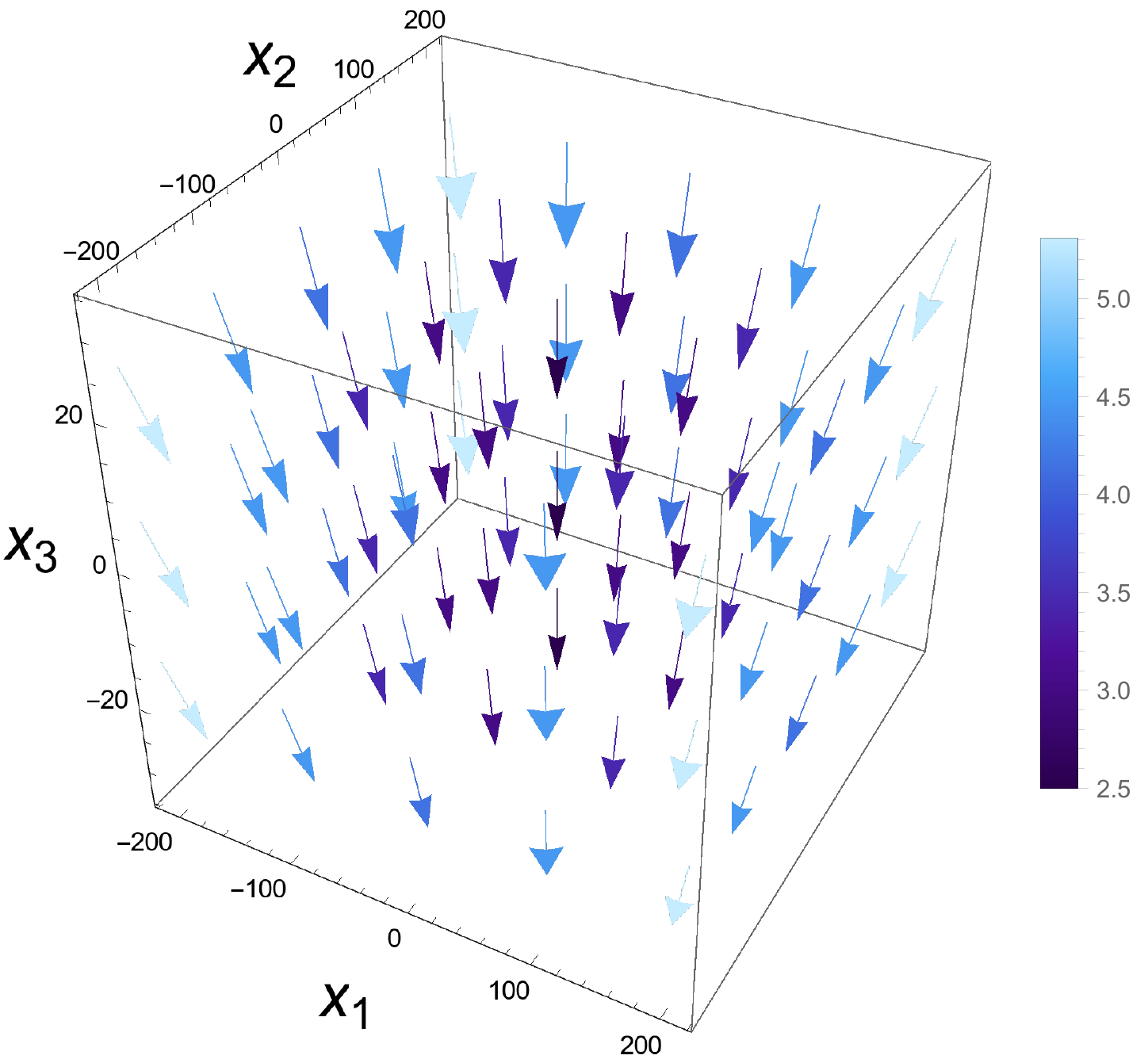}
\endminipage\hfill
\minipage{0.32\textwidth}%
  \includegraphics[width=4.75cm]{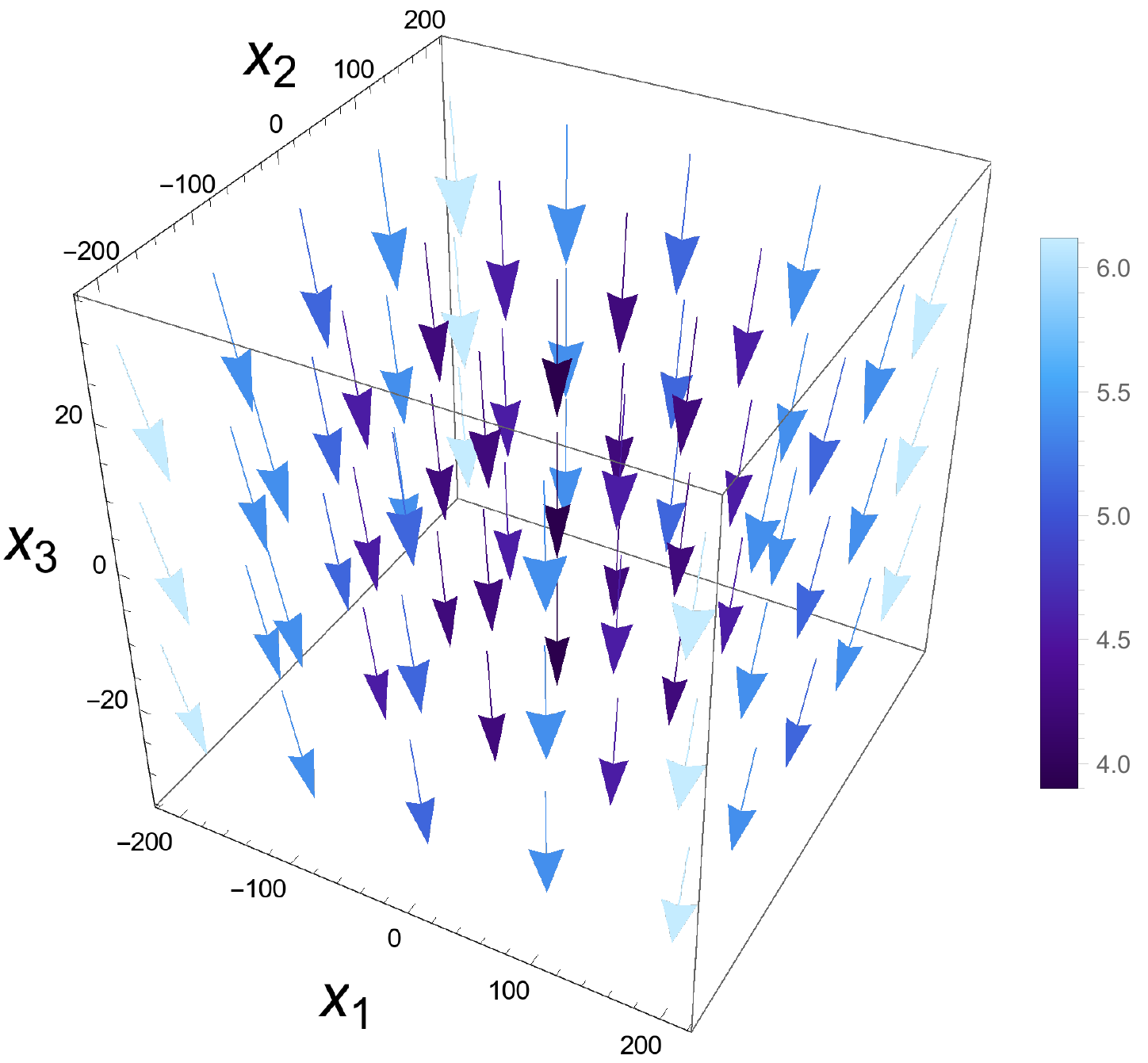}
\endminipage
  \caption{Time evolution of the one point function of the current $\braket{\vec{J}}$ is displayed in three time slices, from left to right,  $t=0$, $t=1.87445$, $t=3.74976$. The vectors are shaded according to $|\vec{J}(t)|$. \textit{Left:} The current is initially directed radially inward toward the $x_3$-axis. 
  \textit{Mid:} As the total charge increases and is accelerated by the electric field the current flow is closer to being directed entirely along $x_3$. \textit{Right:} Near the end of the simulated window the total charge has increased significantly, near the axis the contribution of the current in the transverse plane is dwarfed by the contribution in the $x_3$-direction along the aligned electric and magnetic fields. 
    \label{fig:CurrentEvolution}}
\end{figure}

\begin{figure}[!htb]
\minipage{0.32\textwidth}
  \includegraphics[width=4.75cm]{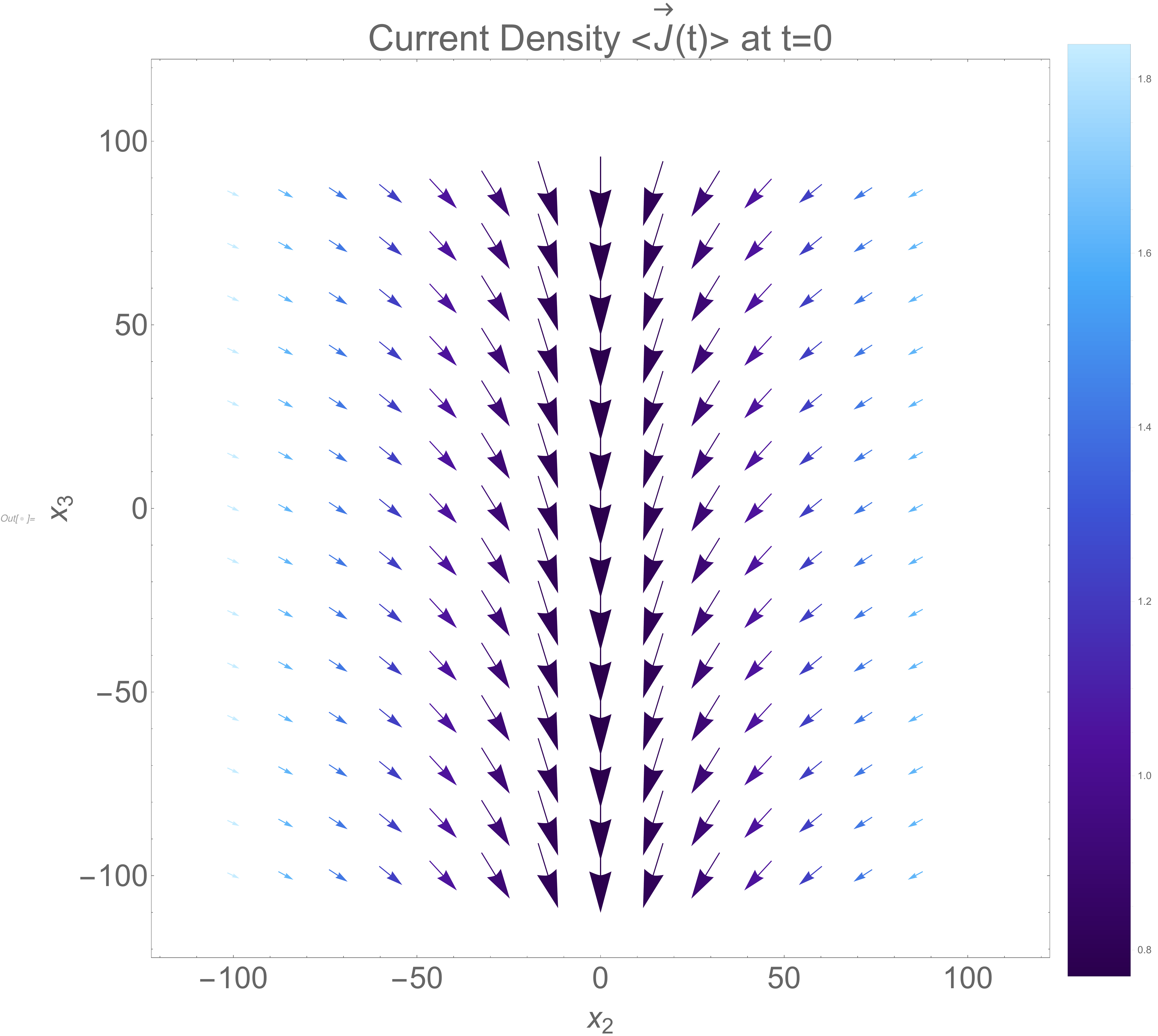}
\endminipage\hfill
\minipage{0.32\textwidth}
  \includegraphics[width=4.75cm]{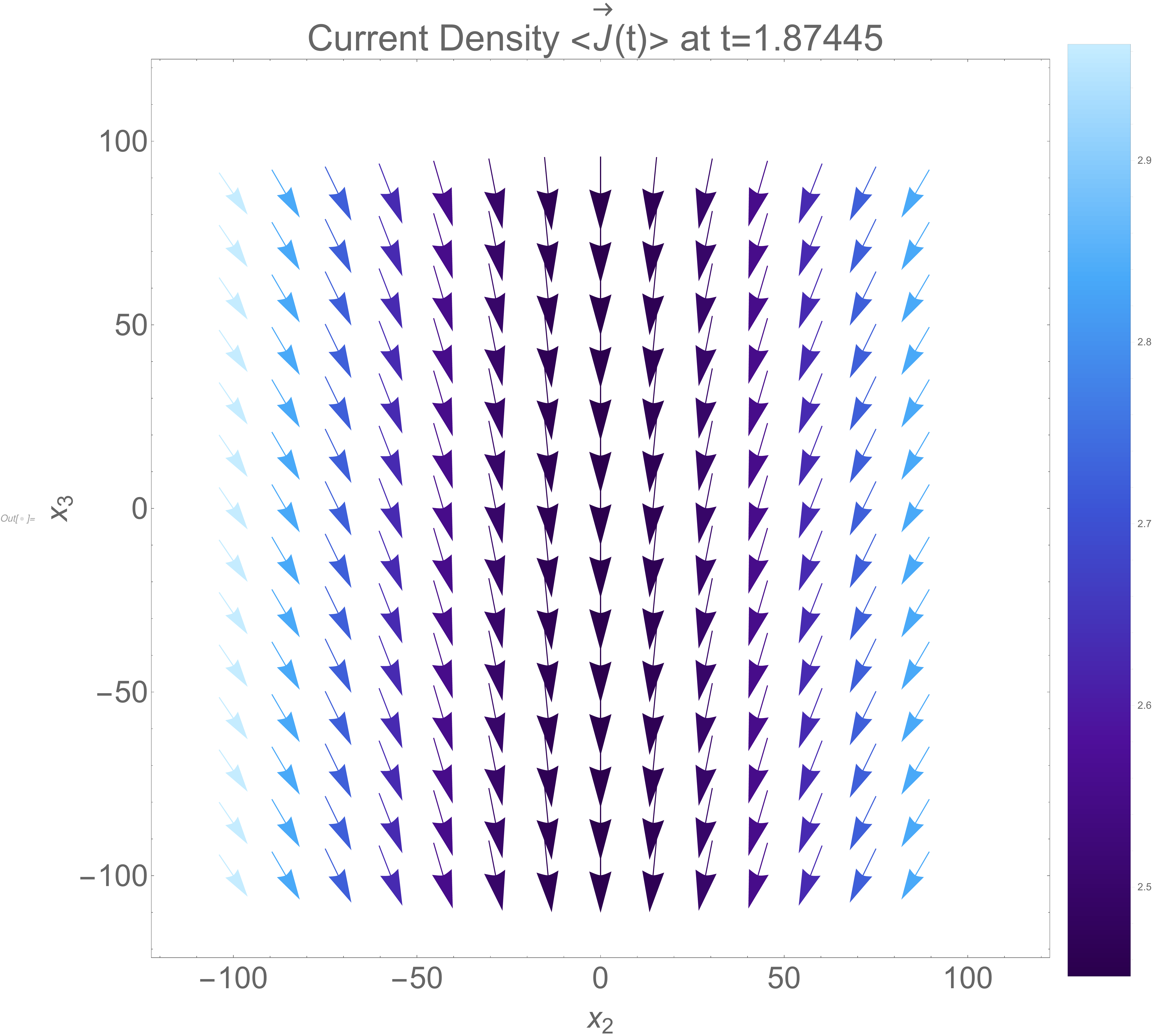}
\endminipage\hfill
\minipage{0.32\textwidth}%
  \includegraphics[width=4.75cm]{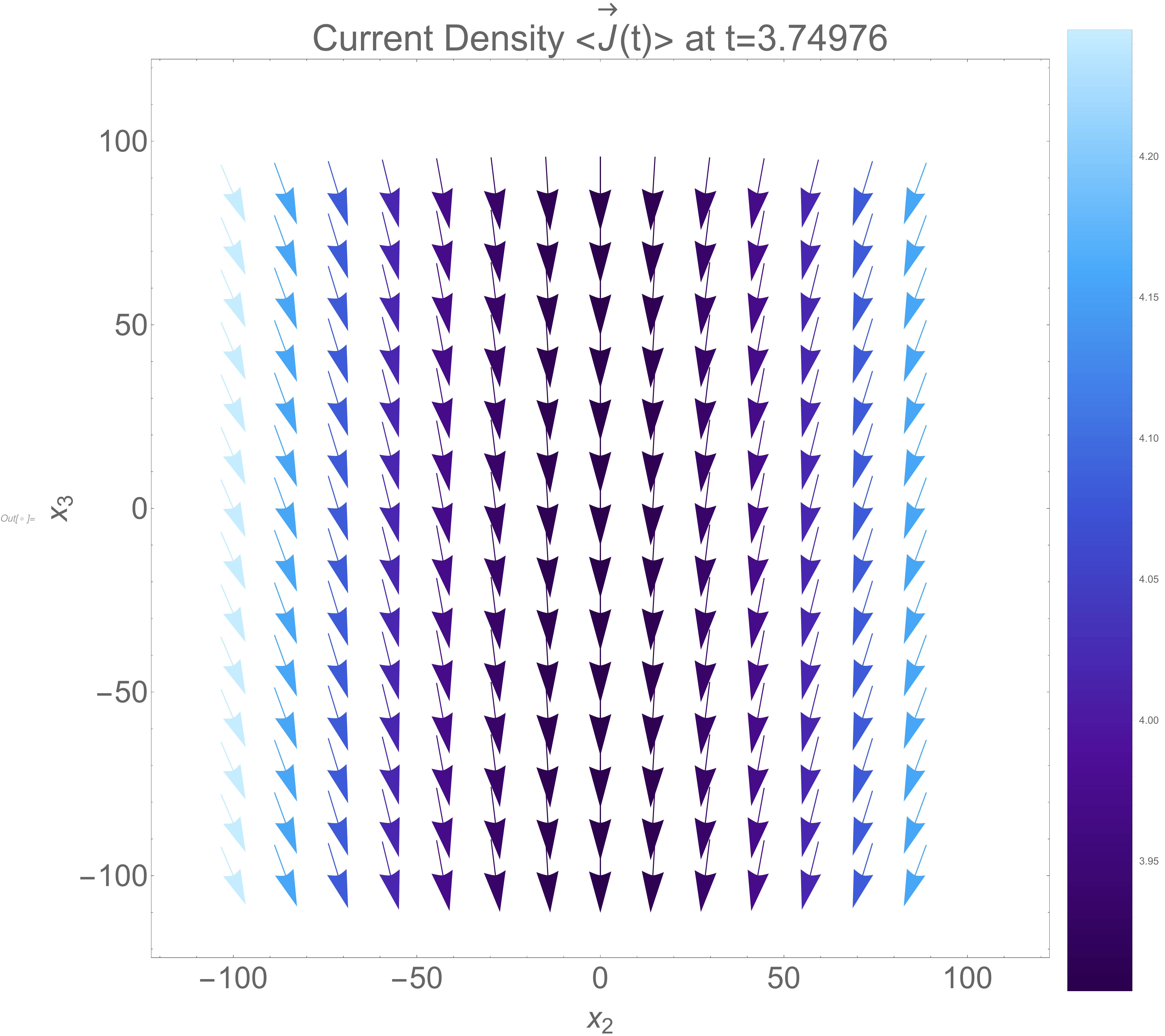}
\endminipage
  \caption{Time evolution of the one point function of the current $\braket{\vec{J}}$ is displayed in three time slices, from left to right,  $t=0$, $t=1.87445$, $t=3.74976$ at $x_1=0$. The vectors are shaded according to $|\vec{J}(t)|$. \textit{Left:} The current is initially directed radially inward toward the $x_3$-axis. 
  \textit{Mid:} As the total charge increases and is accelerated by the electric field the current flow is closer to being directed entirely along $x_3$. \textit{Right:} Near the end of the simulated window the total charge has increased significantly, near the axis the contribution of the current in the transverse plane is dwarfed by the contribution in the $x_3$-direction along the aligned electric and magnetic fields. 
    \label{fig:xycutCurrent}}
\end{figure}

In figure~\ref{fig:CurrentEvolution} we display the evolution of the spatial components $\braket{J^i(t)}$ of the dual current. We display this vector field at three different times during the evolution of the plasma. The left image of figure~\ref{fig:CurrentEvolution} is taken when the system begins its evolution at $t=0$. We can see that we have an azimuthally symmetric flow of axial charge directed approximately towards the $x_3$-axis. We can see the beginning of a flow of this current in the $x_3$-direction with the vectors all pointing slightly down along the $x_3$. The middle image of figure~\ref{fig:CurrentEvolution} displays the current $\braket{J^i(t)}$ at approximately half way through the evolution with $t=1.87445$. In this image we can continue to see the current flowing in towards the $x_3$-axis. However we also see a more significant change in the orientation of the vector field. At this point in the evolution it is clear the flow is directed along the $x_3$-axis. In the right image of figure~\ref{fig:CurrentEvolution} we are near the end of the simulation window at $t=3.74976$. At this point in the evolution the flow within the spatial window displayed is almost entirely directed in along the $x_3$-axis. It is interesting to note that if we choose our window to include a larger spatial extent we would see an image similar to the left image of figure~\ref{fig:CurrentEvolution}. Within a spatial range of $(x_1,x_2)\in(-200,200)\times(-200,200)$ the current is directed almost entirely along the $x_3$-axis at the late times in our evolution. However outside this range the vectors asymptote to an azimuthally symmetric radially inflowing current. The same three time slices are displayed in figure~\ref{fig:xycutCurrent} plotted in the $x_2-x_3$ plane at $x_1=0$. \\

\noindent\textbf{Hydrodynamics:} The behavior of the heat current and the charge density can be understood by considering a simple hydrodynamic model. Consider a fluid of axial charges coupled to an external electromagnetic field. The equations of motion are the conservation equations~\cite{Son:2009tf},
\begin{align}
 \partial_{\mu} T^{\mu\nu}&=F^{\nu\lambda}J_{\lambda}, \\
 \partial_{\mu}J^{\mu} &= -\frac{ C}{8} \epsilon^{\alpha\beta\gamma\delta}F_{\alpha\beta}F_{\gamma\delta}.
\end{align}
We take an energy momentum tensor of the form,
\begin{equation}
 T^{\mu\nu}=\epsilon \delta^{\mu}_{0}\delta^{\nu}_0+ J_h\left( \delta^{\mu}_{0}\delta^{\nu}_3+ \delta^{\mu}_{3}\delta^{\nu}_0\right)+ p_t \delta^{\mu}_{i} \delta^{\mu}_{j}+p_l\delta^{\mu}_3\delta^{\nu}_3,
\end{equation}
with $i,j=1,2$ and the energy density $\epsilon$, heat current $J_h$, transverse pressure $p_t$ and longitudinal pressure $ p_l$ are functions of time only. We take the current to have the form,
\begin{equation}
 J^{\mu}=(J^0,-Cx_1 B E,-Cx_2 B E,J^3),
\end{equation}
with $J^0$ the charge density and $J^3$ the current density in the $x_3$ direction to be functions of time only. The conservation of energy-momentum reduces to two equations, taken together with the conservation of current forms a system of first order ODE's
\begin{align}
\partial_t J^0&=C E_3B_3, \label{eq:charge}\\
\partial_t J_h&=E_3 J^0, \label{eq:heat}\\
 \partial_t \epsilon &= E_3 J^3 \, . \label{eq:energy} 
\end{align}
We first solve eq.\ (\ref{eq:charge}) for $J^0(t)$ whose solution can be substituted into eq.\ (\ref{eq:heat}) to obtain $J_h(t)$ leading to,
\begin{align}
 J^0(t)&=C E_3 B_3 t+ q_0, \label{eqn:hydrochargecurrent}\\
 J_h(t)&= E_3\left(C E_3 B_3 \frac{t^2}{2}+q_0t\right)+J_{h0} \, .\label{eqn:hydroheatcurrent}
\end{align}
We find the solutions given in eq.\ (\ref{eqn:hydrochargecurrent}) and eq.\ (\ref{eqn:hydroheatcurrent}) are exactly the relations we found from a near boundary solution to the Einstein equations. As is typical in holographic systems we find hydrodynamics contained within the bulk gravitational evolution. The dynamics
of the axial charge density and the heat current for our setup follow exactly as predicted by hydrodynamics. Furthermore the final hydrodynamic equation, eq.\ (\ref{eq:energy}), displays the rate of change of the energy density to be given by the Joule heating term found in section~\ref{sec:Asymptotic}. It should be noted that, although powerful, hydrodynamics was not able to provide for us the behavior of the current $J^3$. The behavior of this quantity must be obtained from another source. In this work we obtained its behavior from the full evolution of the bulk geometry. \\

 \noindent\textbf{Entropy Production: }As an application we consider the entropy produced during the process of isotropization. A standard definition of the out of equilibrium thermal entropy is given by the area of the apparent horizon. However this definition is not unique, there are many notions of entropy for spacetimes undergoing dynamical processes along with many area increase laws~\cite{Bhattacharyya:2008xc,Lewkowycz:2013nqa,Bousso:2015mqa,Sanches:2016pga}. 

\begin{figure}[!htb]
\minipage{0.48\textwidth}
  \includegraphics[width=7.25cm]{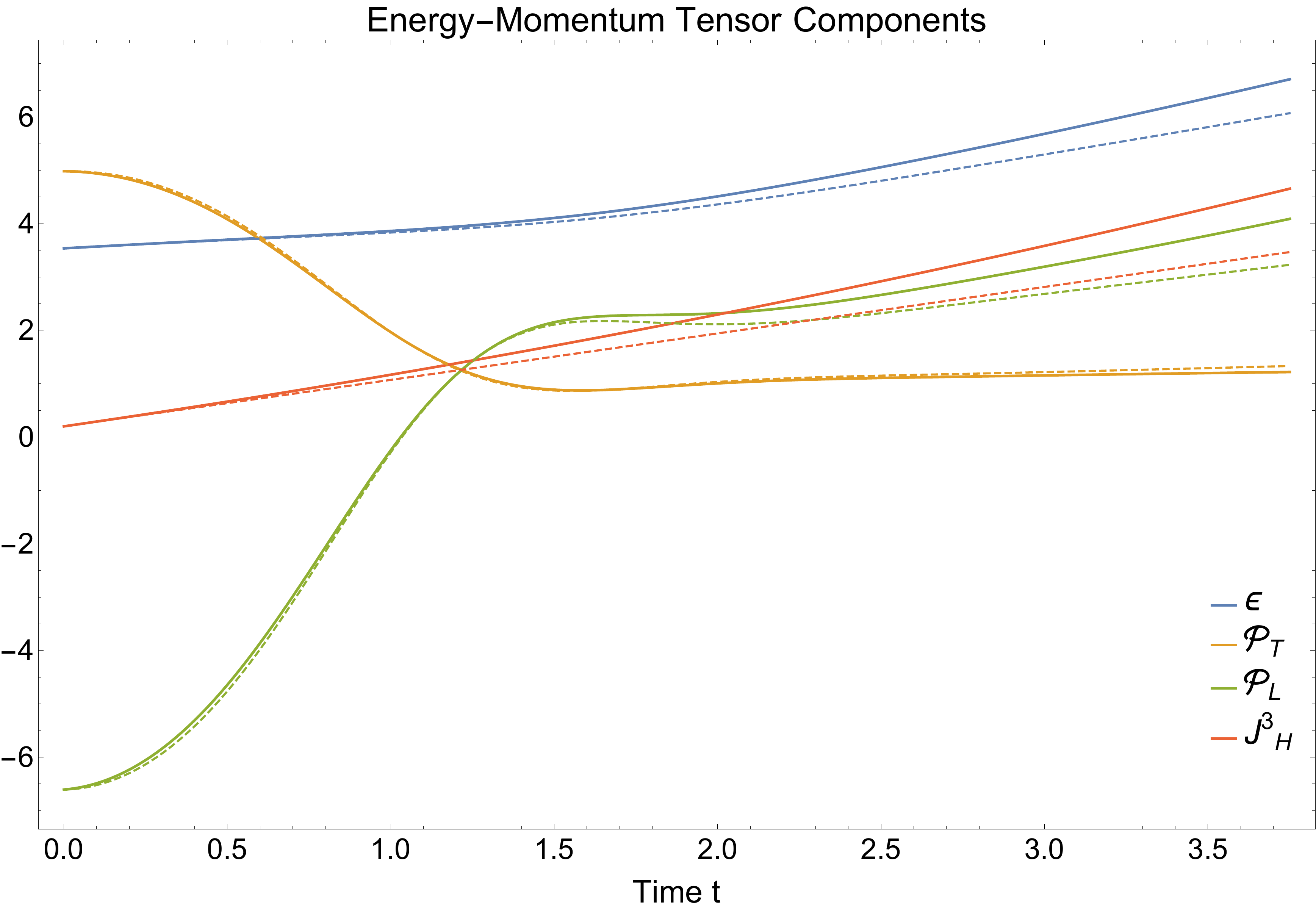}
\endminipage\hfill
\minipage{0.48\textwidth}
  \includegraphics[width=7.25cm]{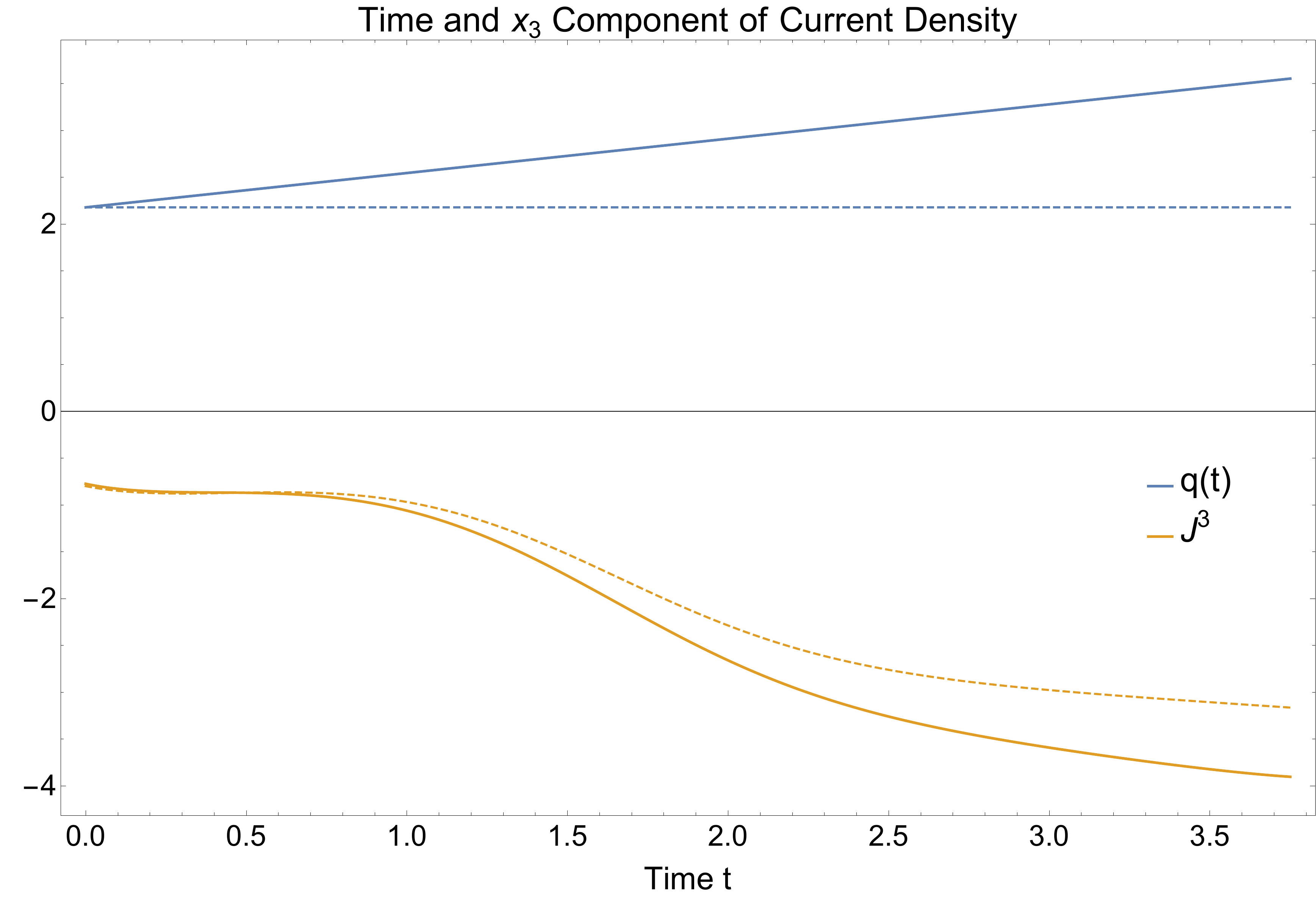}
\endminipage\hfill
  \caption{Time evolution of a strongly coupled far from equilibrium plasma with an axial anomaly. In both images the dashed lines are the evolution with the Chern-Simons coupling $\gamma=0$. \textit{Left:} The time evolution of the one point functions of the energy-momentum tensor are displayed at $\mu_r=1$. The blue line is the energy density $\epsilon=\braket{T_{00}}$, the green line the longitudinal pressure $\mathscr{P}_L=\braket{T_{33}}$, the orange line the transverse pressure $\mathscr{P}_T=\frac{1}{2}(\braket{T_{11}}+\braket{T_{22}})$ and the red line is the heat current $J^3_H=\braket{T_{03}}$. \textit{Right:} The time evolution of the time dependent components of the current density $\braket{J^{\mu}}$ are displayed. The blue line is the axial charge density $\braket{J^{0}(t)}$ and the orange line is the $x_3$ component of the $\braket{J^{3}(t)}$.  
    \label{fig:gamma_0}}
\end{figure}

Keeping in mind eq.\ (\ref{eqn:units}) the entropy density can be calculated via the spatial scale factor~\cite{wilkethesis},
\begin{equation}
    s(t)=4\pi S(t,z_h(t))^3,\label{eqn:Apparent_Entropy}
\end{equation}
although it should be stated that only near equilibrium can we truly call this quantity the entropy density in the dual theory. In order to put in context the generation of entropy during the production of axial charges we choose to compare our data to the same setup only with the Chern-Simons coupling $\gamma=0$. In figure~\ref{fig:gamma_0} we compare the results of evolving our system with and without the Chern-Simons coupling. The dashed lines represent the evolution with $\gamma=0$. In the left image of figure~\ref{fig:gamma_0} we can see that without the Chern-Simons coupling we have a decrease in the growth of the energy density. This is due to a decrease in the current density component $\braket{J^3}$ as can be seen in the right image of figure~\ref{fig:gamma_0}. Accompanying this curve we also see that we have a fixed charge density throughout the evolution as without the Chern-Simons coupling there is no anomalous production of charges. The difference in the evolution of the energy density leads to changes in the evolution of the pressures while the transverse pressure is roughly the same the longitudinal pressure is decreased.

\begin{figure}[!htb]
\minipage{0.48\textwidth}
  \includegraphics[width=7.25cm]{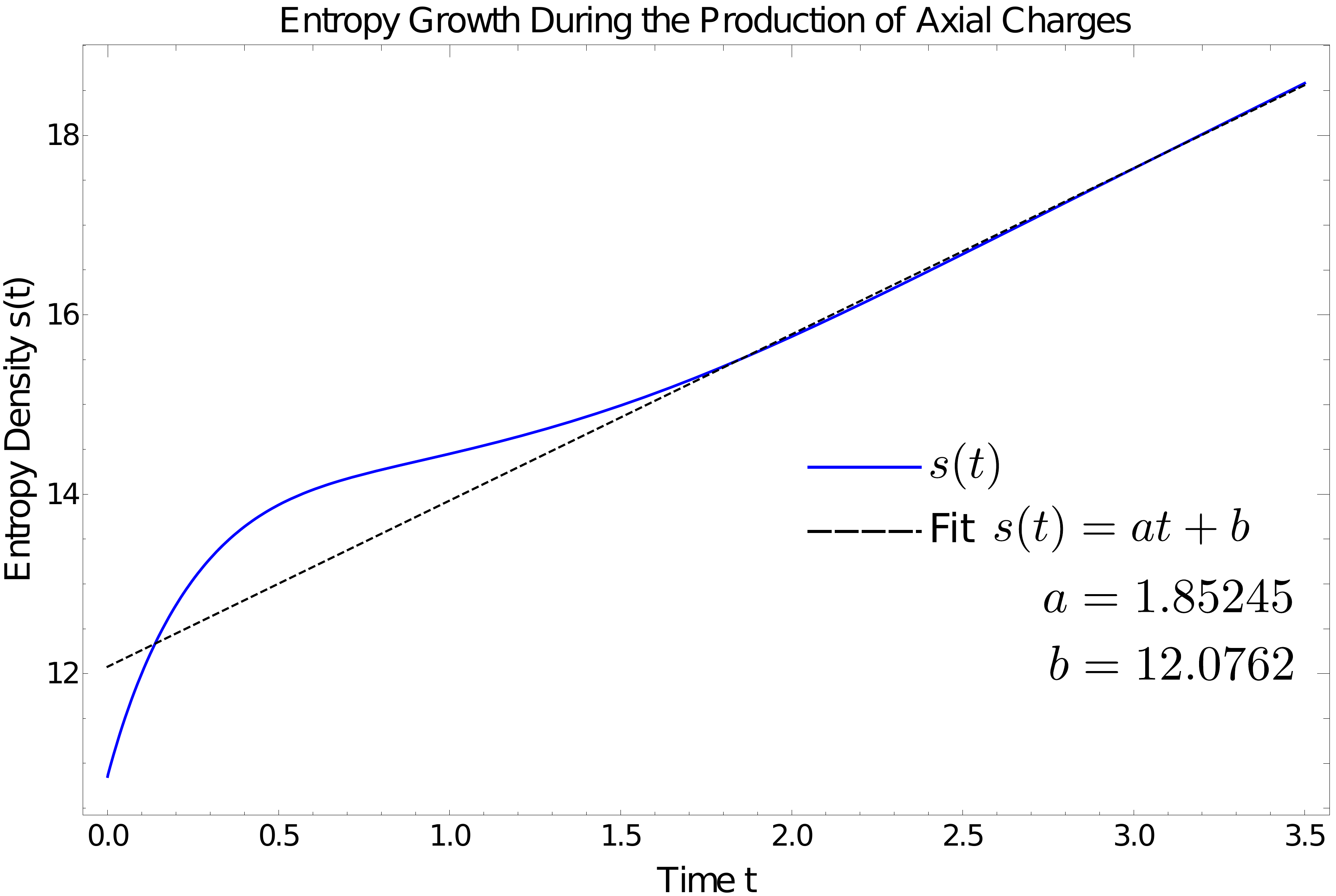}
\endminipage\hfill
\minipage{0.48\textwidth}
  \includegraphics[width=7.25cm]{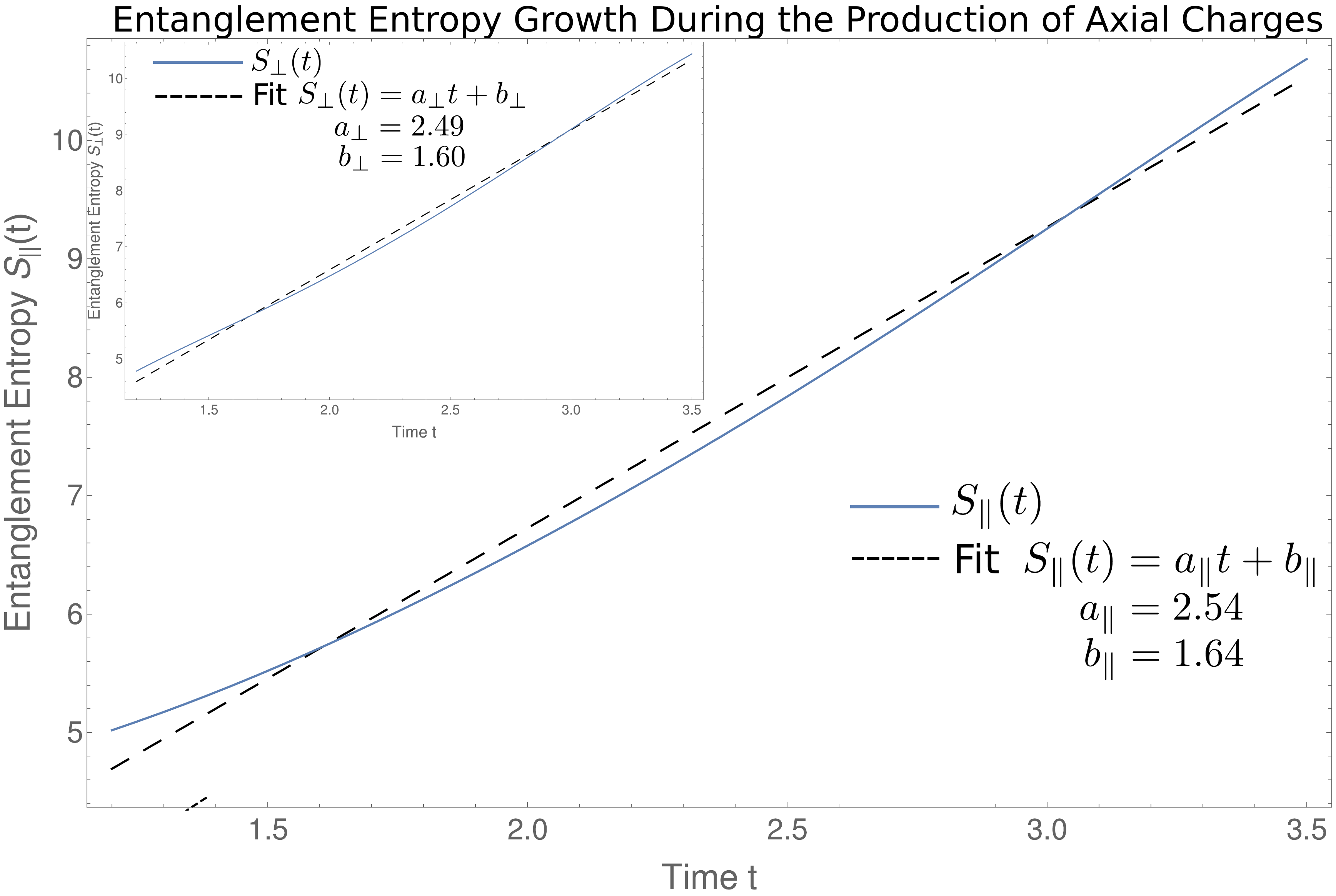}
\endminipage\hfill
  \caption{
  \textit{Left:} Time evolution of the entropy density $s(t)$ is shown in the figure by the blue line. The dashed black line is a linear fit of this data near late times. One can see slight oscillations of the blue curve around this line. \textit{Right:} The evolution of the entanglement entropy for a strip like topology with embedding coordinates $(v(\sigma),z(\sigma),x_3(\s))$ is displayed in the figure by the blue curve. The dashed black line represents a linear fit to this data near the late times. The inset displays the same information of the evolution of the entanglement entropy for a strip like topology but with embedding coordinates $(v(\sigma),z(\sigma),x_1(\s))$. In both cases the entangling region had a width of $\ell=0.8$.
    \label{fig:Entropy_With_Axial}}
\end{figure}

In figure~\ref{fig:Entropy_With_Axial} we display both the thermal and entanglement entropy produced during isotropization of the plasma with aligned electric and magnetic fields. We can see in the left image of figure~\ref{fig:Entropy_With_Axial} the growth of entropy in the system is a monotonic function of time. After a sufficiently long time the function approaches a linear growth. Displayed in the figure is a fit to this linear growth with a growth rate of $\exd s/\exd t=1.85245$. The linear growth of the thermal entropy in the evolution of SYM plasma is not a new phenomenon it was recently seen and discussed in the context of phenomenological insights gained from holographic heavy ion collisions~\cite{Muller:2020ziz} (see their work for more information). It should be noted that the linear growth of the entropy as displayed in~\cite{Muller:2020ziz} occurs before thermalization and without sourcing\footnote{We thank the referee for pointing this out.}. Hence the behavior we observe is by definition of different origin as will be discussed further in this section. 

In the right image of figure~\ref{fig:Entropy_With_Axial} we display the growth of the entanglement entropy in both the transverse and longitudinal directions. We see that the entanglement entropy oscillates weakly around a linear growth in time. The linear growth of the entanglement entropy is a familiar feature of systems undergoing a global quench~\cite{2005JSMTE04010C} (see also~\cite{AbajoArrastia:2010yt,Balasubramanian:2011ur} for early examples in holography). We also display a fit to this data with the rate of growth of the entanglement entropy in the transverse and longitudinal directions,
\begin{equation}
    \frac{\exd S_{\perp}}{\exd t}=2.49 \hspace{2cm}
    \frac{\exd S_{\parallel}}{\exd t}=2.54 \, .\end{equation}
It is interesting to note that although the entanglement entropy grows at a slightly faster rate in the longitudinal direction they both grow roughly at a roughly equal rate. The rate of growth of the entanglement entropy during the linear regime is proportional to the entanglement velocity. This linear regime is what is referred to as the post-local-equilibration regime in~\cite{Liu:2013iza}. In this regime $S(t)=Av_Es_{eq}t$ with $A$ the area of the region and $s_{eq}$ the value of the entropy density of the equilibrium state. It is however unclear what equilibrium state we should compare to. 

\begin{figure}[!htb]
\minipage{0.48\textwidth}
  \includegraphics[width=7.25cm]{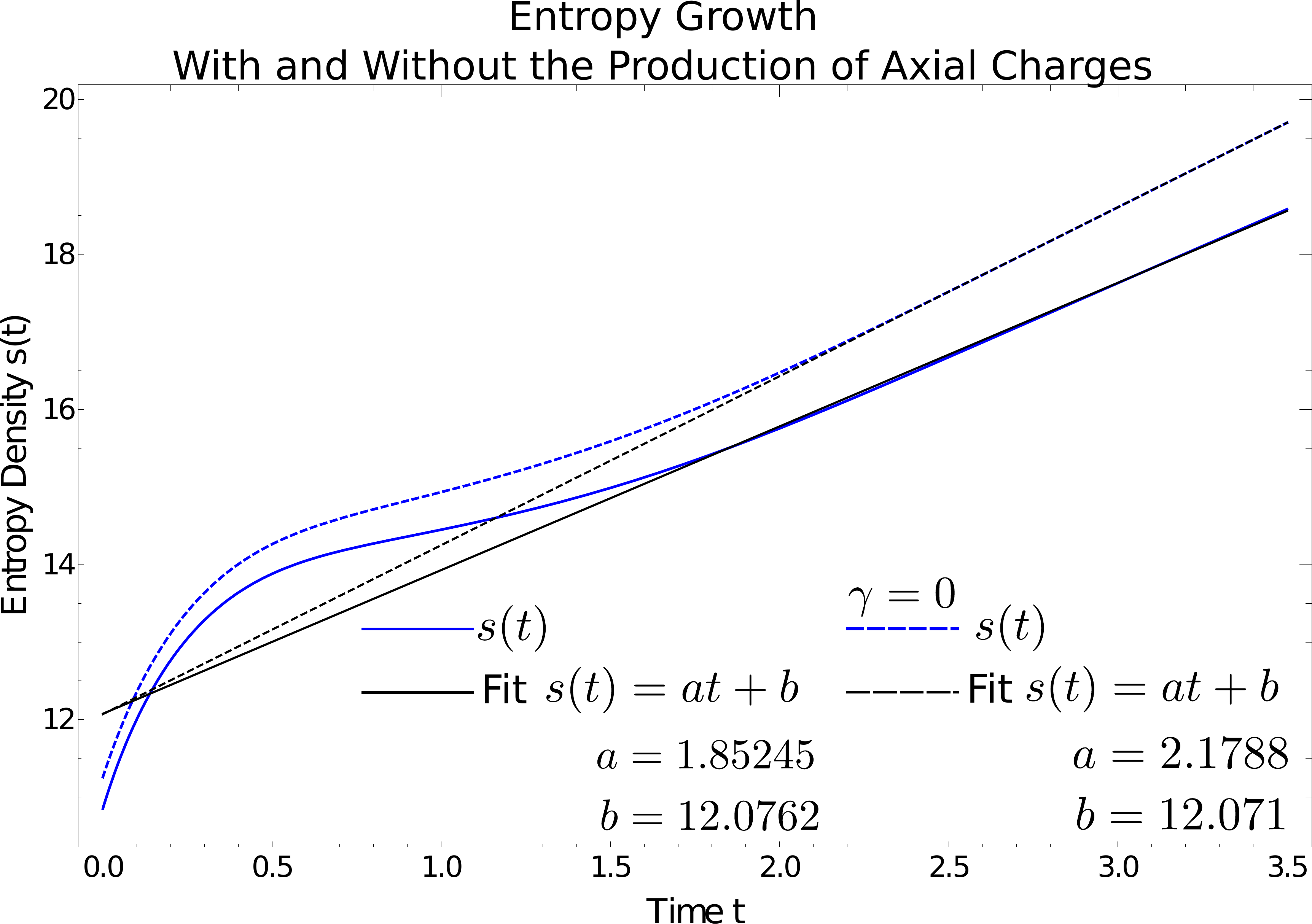}
\endminipage\hfill
\minipage{0.48\textwidth}
  \includegraphics[width=7.25cm]{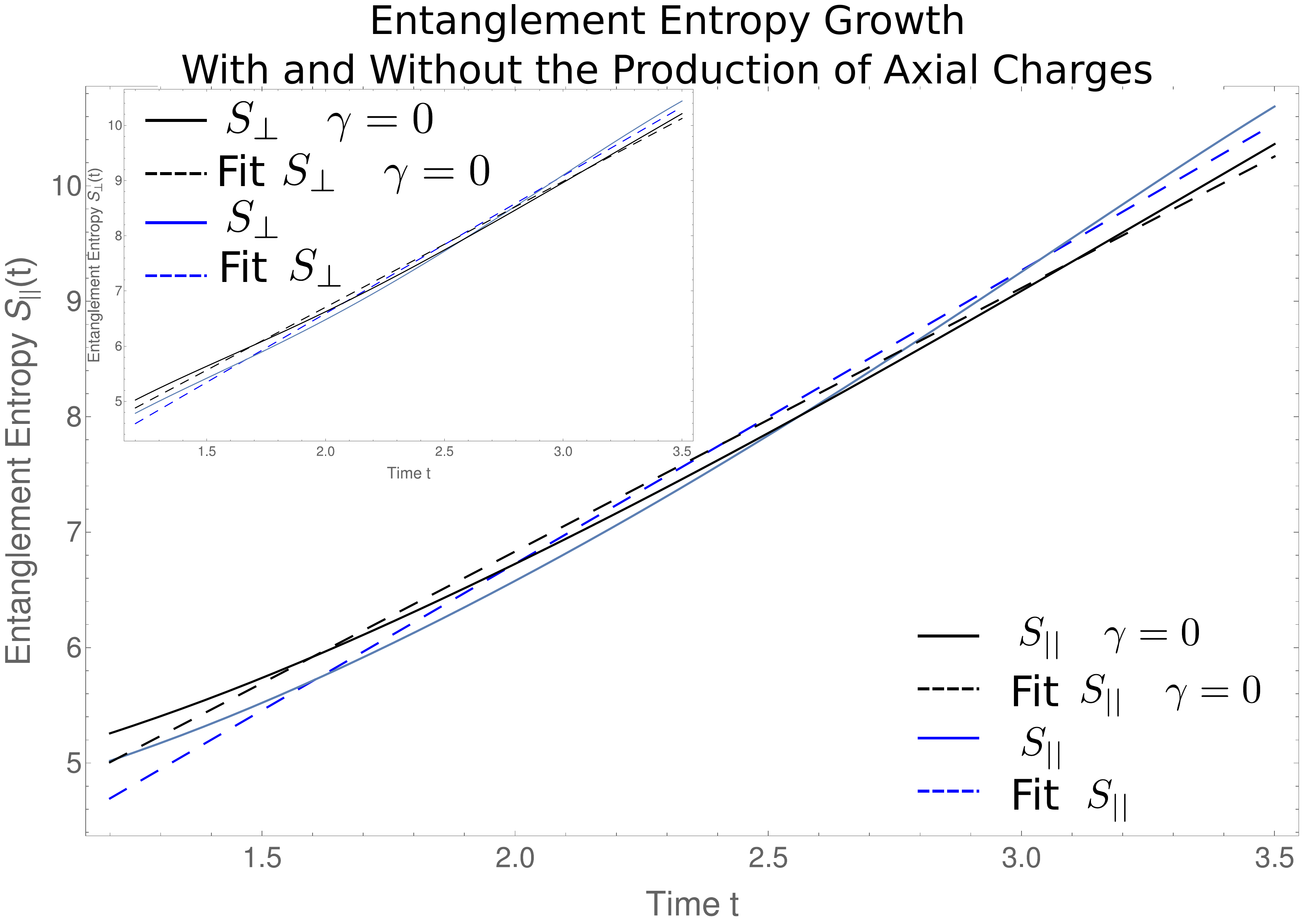}
\endminipage\hfill
  \caption{\textit{Left:} Time evolution of entropy is displayed with solid lines representing $\gamma=1/2$ and dashed lines representing $\gamma=0$.
 The evolution of the entropy density $s(t)$ is shown in the figure by blue curves. While the black lines are linear fits to data near late times. Fit parameters are displayed in the plot. \textit{Right:} The evolution of the entanglement entropy for a strip like topology with embedding coordinates $(v(\sigma),z(\sigma),x_3(\s))$ is displayed in the figure with blue lines representing $\gamma=1/2$ and black lines representing $\gamma=0$. Dashed lines represent linear fits to this data near the late times. The inset displays the same information of the evolution of the entanglement entropy for a strip like topology but with embedding coordinates $(v(\sigma),z(\sigma),x_1(\s))$. In both cases the entangling region had a width of $\ell=0.8$. To avoid unnecessary clutter the fit parameters displayed in table~\ref{tab:FitParameters} are not displayed on the plot. 
    \label{fig:Entropy_comparison}}
\end{figure}

In figure~\ref{fig:Entropy_comparison} we display the evolution of the entropy and the entanglement entropy with and without the production of chiral charges. We compute this in both the direction parallel and transverse to the aligned electric and magnetic field. In the left image of figure~\ref{fig:Entropy_comparison} we display the entropy during the production of axial charges as a solid line and without the production of axial charges as a dashed line. We can see in the left image of figure~\ref{fig:Entropy_comparison} that turning on the Chern-Simons coupling leads to smaller growth rate of the entropy,
\begin{equation}
   \frac{\exd s}{\exd t}< \frac{\exd s_{\gamma=0 }}{\exd t} .
\end{equation}
In the right image of figure~\ref{fig:Entropy_comparison} we display the entanglement entropy during the production of axial charges as solid blue lines and without the production of axial charges as solid black lines. We provide the linear fits to all of these curves in the plot to help guide the eye towards the late time linear regime. The colors of the dashed fit lines are in correspondence with colors of the solid lines. We can see in both the transverse and longitudinal direction that although the entanglement entropy is larger at earlier times when the Chern-Simons coupling is turned off, it has a smaller growth rate (see table~\ref{tab:FitParameters}),
\begin{equation}
   \frac{\exd S_{\perp}}{\exd t}> \frac{\exd S_{\perp,\gamma=0 }}{\exd t} \hspace{2cm}
    \frac{\exd S_{\parallel}}{\exd t}>\frac{\exd S_{\parallel,\gamma=0 }}{\exd t}.
\end{equation}
Hence we observe an increased entanglement velocity with a non-zero Chern-Simons coupling. We suspect this increase in the entanglement velocity is related to the azimuthally symmetric inflow of current re-aligning itself to a flow along the $x_3$ axis and the increasing number of axial charges. We also suspect the initially larger value of the entanglement entropy without a Chern-Simons coupling is due to an already aligned current flowing along the $x_3$ axis. 

\begin{table}
\begin{center}
\begin{tabular}{c|c|c|c|c}
  & $a_{\perp}$ & $b_{\perp}$ & $a_{\parallel}$ & $b_{\parallel}$   \\
  \hline 
 $\gamma=0$    & $2.28$ & $2.15$ &$2.28$ &$2.27$ \\
 $\gamma=1/2$  & $2.49$   & $1.60$ &$2.54$ &$1.64$ 
\end{tabular}
\end{center}
\caption{\label{tab:FitParameters}We display the parameters found by fitting the late time evolution of the entanglement entropy to a linear curve of the form $S_{\perp,\parallel}=a_{\perp,\parallel}t+b_{\perp,\parallel}$. We fit this data for both $\gamma=0$ and $\gamma=1/2$ while holding fixed all other parameters. }
\end{table}

\begin{figure}
 \minipage{0.38\textwidth}
  \includegraphics[width=6cm]{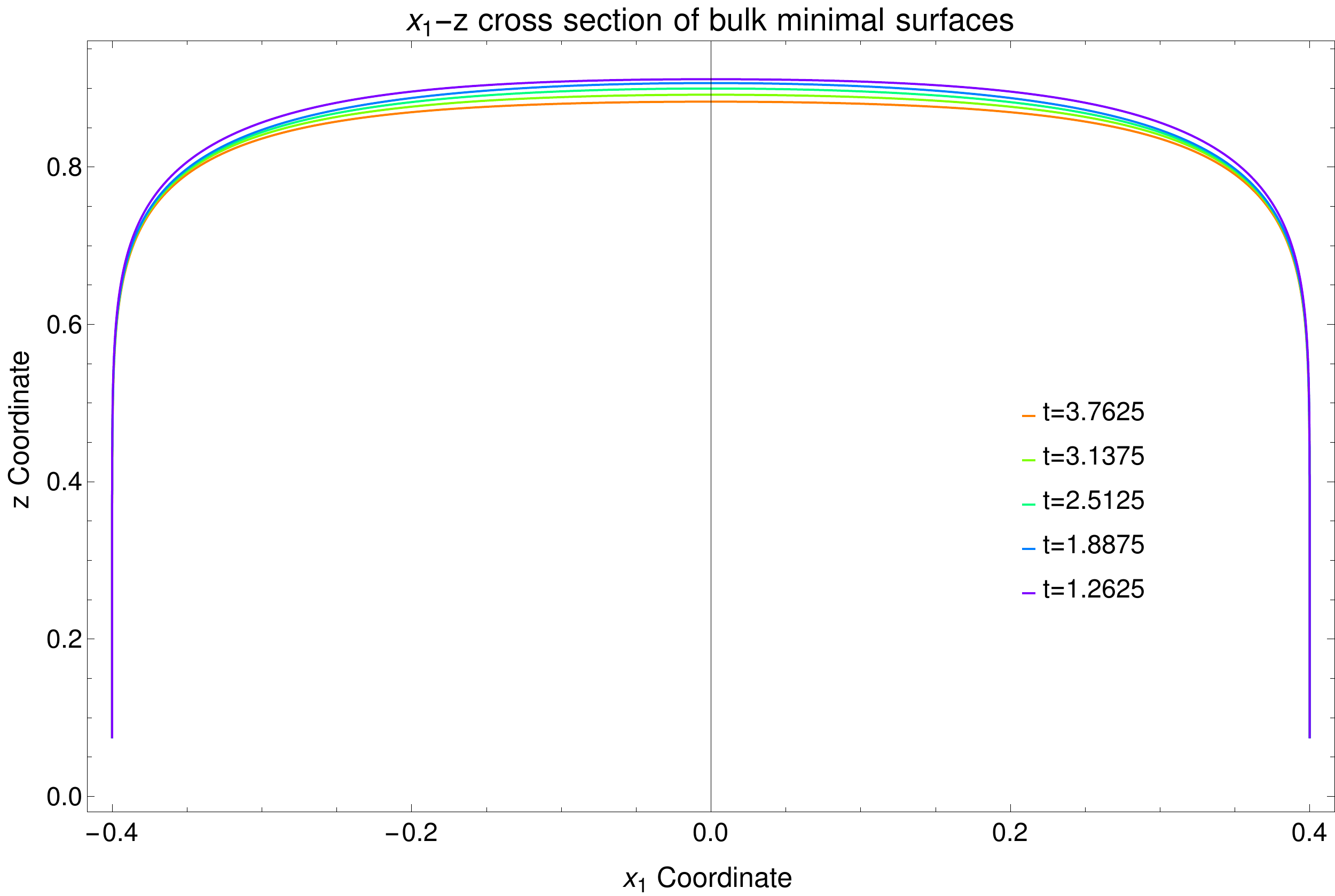}
\endminipage\hfill
\minipage{0.55\textwidth}
  \includegraphics[width=8.8cm]{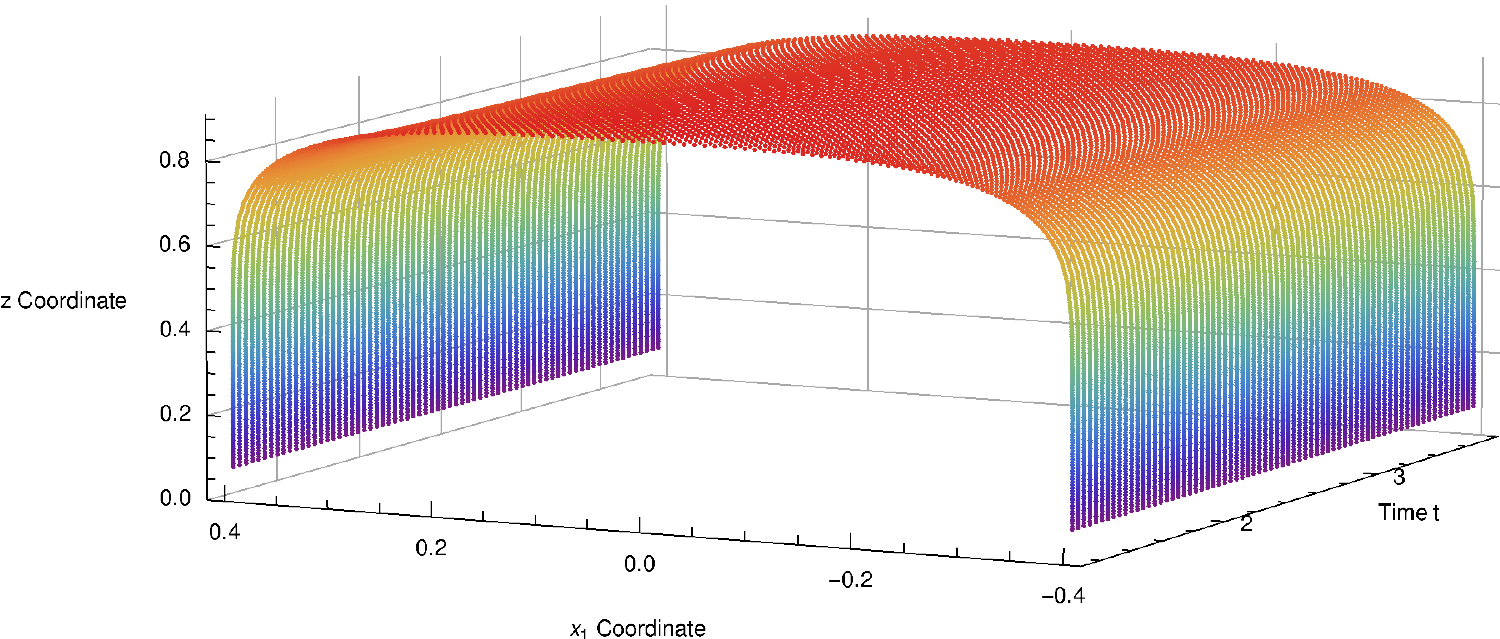}
\endminipage\hfill
\caption{\label{fig:surfaces}\textit{Left:} Cross sections of the minimal surfaces calculated in the geometry dual to the time evolution of a strongly coupled far from equilibrium plasma with an axial anomaly. The surfaces displayed were calculated in the transverse direction with embedding coordinates $(v(\sigma),z(\sigma),x_1(\sigma))$. In the image we display $(x_1(\sigma),z(\sigma))$ at various times during the evolution. We can see the surfaces penetrate progressively deeper into the bulk geometry as time goes on. \textit{Right:} The bulk minimal surfaces in the transverse direction calculated in the geometry dual to the time evolution of a strongly coupled far from equilibrium plasma with an axial anomaly.}
\end{figure}
\begin{figure}
 \minipage{0.52\textwidth}
  \includegraphics[width=7.5cm,height=5.7cm]{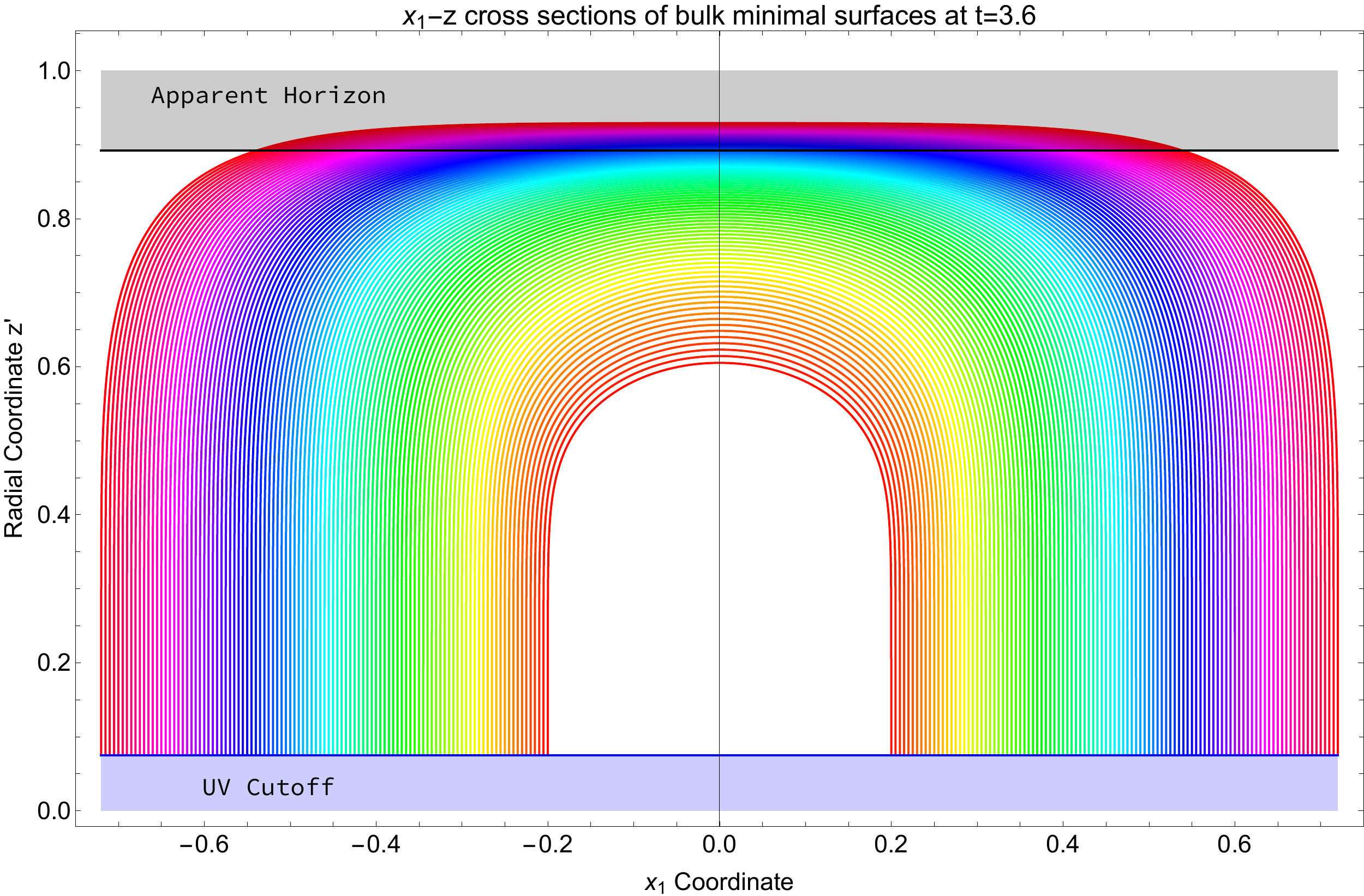}
\endminipage\hfill
\minipage{0.52\textwidth}
  \includegraphics[width=7.5cm]{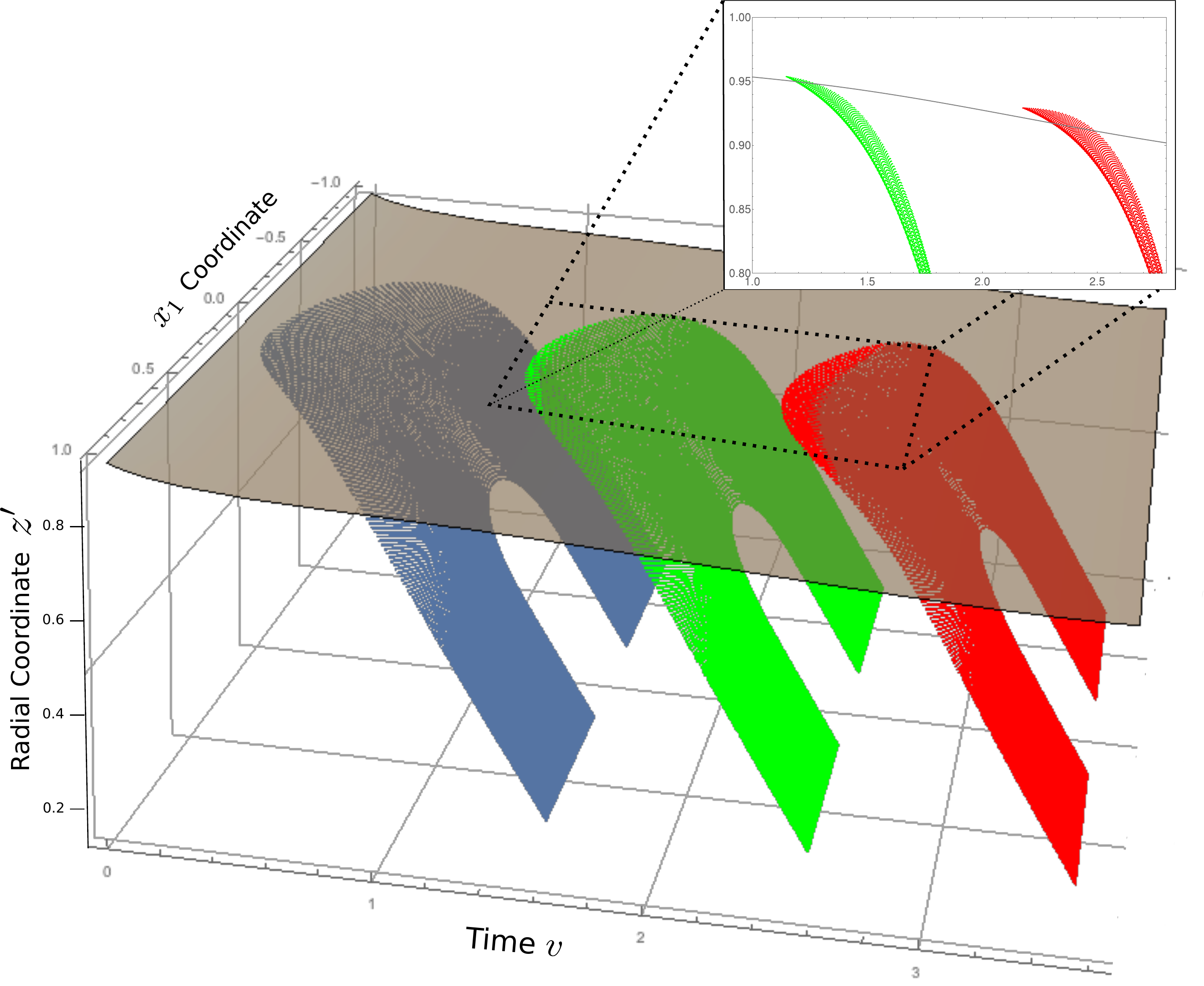}
\endminipage\hfill
\caption{\label{fig:surfaces_Length_dep}Minimal surfaces calculated in the geometry dual to the time evolution of a strongly coupled far from equilibrium plasma with an axial anomaly. The displayed surfaces were calculated in the transverse direction for strips of width $\ell\in [0.4,1.45]$. \textit{Left:} Cross sections of the surfaces in the $x_1-z'$ plane where $z'$ is the original un-shifted z coordinate at time $t=3.6$. We display the current location of the apparent horizon as a solid black line and shade the region behind the horizon. We display the cutoff surface $z_{uv}=0.075$ as a blue line and shade the region between the cutoff and the conformal boundary. \textit{Right:} We display three families of bulk minimal surfaces for $\ell\in [0.4,1.45]$ at the times $t=1.6$ (blue), $t=2.6$ (green) and $t=3.6$ (red). The gray surface indicates the location of the apparent horizon. The inset displays a $v-z'$ cross section of the curves at $t=2.6$ (green) and $3.6$ (red) with a black line indicating the location of the apparent horizon. }
\end{figure}

In the right image of figure~\ref{fig:surfaces} we display the surface we compute with a boundary separation of $\ell=0.8$ throughout the evolution with a non-zero Chern-Simons coupling in the transverse direction. On the left of this we display various time slices of the figure on the right in the $x_1-z$ plane. We can see the majority of the minimal surface lays parallel to the apparent horizon. The method we used for this work to compute the geometry is intimately tied to using the final grid point of domain as the location of the apparent horizon. Hence the extent of the domain we evolve ends at the apparent horizon. However we find that increasing the width of the strip in the field theory eventually leads to minimal surfaces which cross the apparent horizon. We display this behavior in figure~\ref{fig:surfaces_Length_dep} in the original coordinate system which we label as $z'$. This coordinate system is obtained by transforming the radial coordinate back to the un-shifted coordinate system. In both images of figure~\ref{fig:surfaces_Length_dep} we display families of minimal surfaces at fixed boundary time for lengths $\ell\in [0.4,1.45]$. In the left image we display a cross section of these surfaces at time $t=3.6$ in the $x_1-z'$ plane. The apparent horizon and bulk cutoff surface are orthogonal to the field theory direction $x_1$ and are displayed as black and blue lines respectively. The area behind the apparent horizon and the area between the cutoff surface and the conformal boundary have been shaded in. We can see that as the width of the strip in the field theory grows we eventually cross the apparent horizon. In the right image of figure~\ref{fig:surfaces_Length_dep} we display three families of minimal surfaces for lengths of $\ell\in [0.4,1.45]$ fixed at a boundary time of $t=1.6$ (blue), $t=2.6$ (green) and $t=3.6$ (red). The location of the apparent horizon is displayed as an opaque gray surface.  The blue surfaces, fixed at the earliest time of the three, do not cross the horizon. However the green and red families of minimal surfaces eventually cross the apparent horizon as the width of the strip is increased. We also include an inset which displays the behavior of the minimal surfaces at $t=2.6$ and $t=3.6$ in the $v-z'$ plane with the black curve displaying the apparent horizon.


The situation of bulk minimal surfaces crossing the horizon can be contrasted with the case of the standard Schwarzschild black brane in equilibrium. If we choose the entangling region to include the entire spacetime in the dual field theory the resulting entanglement entropy will be equal to the entropy~\cite{Nishioka:2009un}. In the bulk gravity theory this corresponds to a minimal surface which coincides with the surface of the horizon. Furthermore in~\cite{Hubeny:2012ry} the author proves that while in equilibrium (static geometries) minimal surfaces used to calculate the entanglement entropy will never cross the horizon\footnote{In equilibrium the apparent horizon will coincide with the event horizon~\cite{Chesler:2008hg}}. While this statement is true for static geometries it has been shown to be false for dynamically evolving spacetimes~\cite{AbajoArrastia:2010yt}.

It is interesting to compare this growth of the entropy as defined by the apparent horizon to the thermodynamical entropy as defined via,
\begin{equation}
 s_{th}=(\epsilon+P)/T \label{eqn:Thermal_Entropy}.
\end{equation}
This requires some notion of a temperature and pressure throughout the evolution. However both of these quantities are only well defined in equilibrium. A standard definition of the out of equilibrium temperature is given by $T=a\epsilon^{1/4}$ with $a$ a constant of proportionality. The energy can be read off from the energy-momentum in eq.\ (\ref{eqn:enmomE}). While the non-equilibrium pressures can be read off from the spatial components of the energy-momentum tensor along the diagonal in eq.\ (\ref{eqn:enmomPt}) and eq.\ (\ref{eqn:enmomPl}). Since our system is anisotropic we take $P=(P_L+P_T)/2$ for the pressure in eq.\ (\ref{eqn:Thermal_Entropy}). With these considerations the thermodynamic entropy can be written as,
\begin{equation}
 s_{th}=\frac{a}{\braket{T^{00}}^{1/4}}\left(\braket{T^{00}}+\frac{1}{2}\left(\frac{1}{2}\left(\braket{T^{11}}+\braket{T^{22}}\right)+\braket{T^{33}}\right)\right).\label{eqn:Thermodynamic}
\end{equation}
The coefficient $a$ should be chosen such that our out of equilibrium temperature matches the standard temperature in equilibrium. However our system never relaxes to an equilibrium configuration. Therefore we must fix the coefficient $a$ in a different way. Plotting eq.\ (\ref{eqn:Thermodynamic}) the resulting thermodynamic entropy is linear at late times with a slope proportional to the coefficient $a$. In order to facilitate a comparison we choose to fix the coefficient $a$ such that the growth rate of the entropy at late times agrees between the thermodynamic and apparent horizon entropy. It is not necessarily the case that two separate measures of entropy should agree during the non-equilibrium evolution of the system. However it is sensible to expect their growth at late times when the system is expected to be a thermal theory would agree. 

\begin{figure}[!htb]
 \begin{center}
  \includegraphics[width=8.5cm]{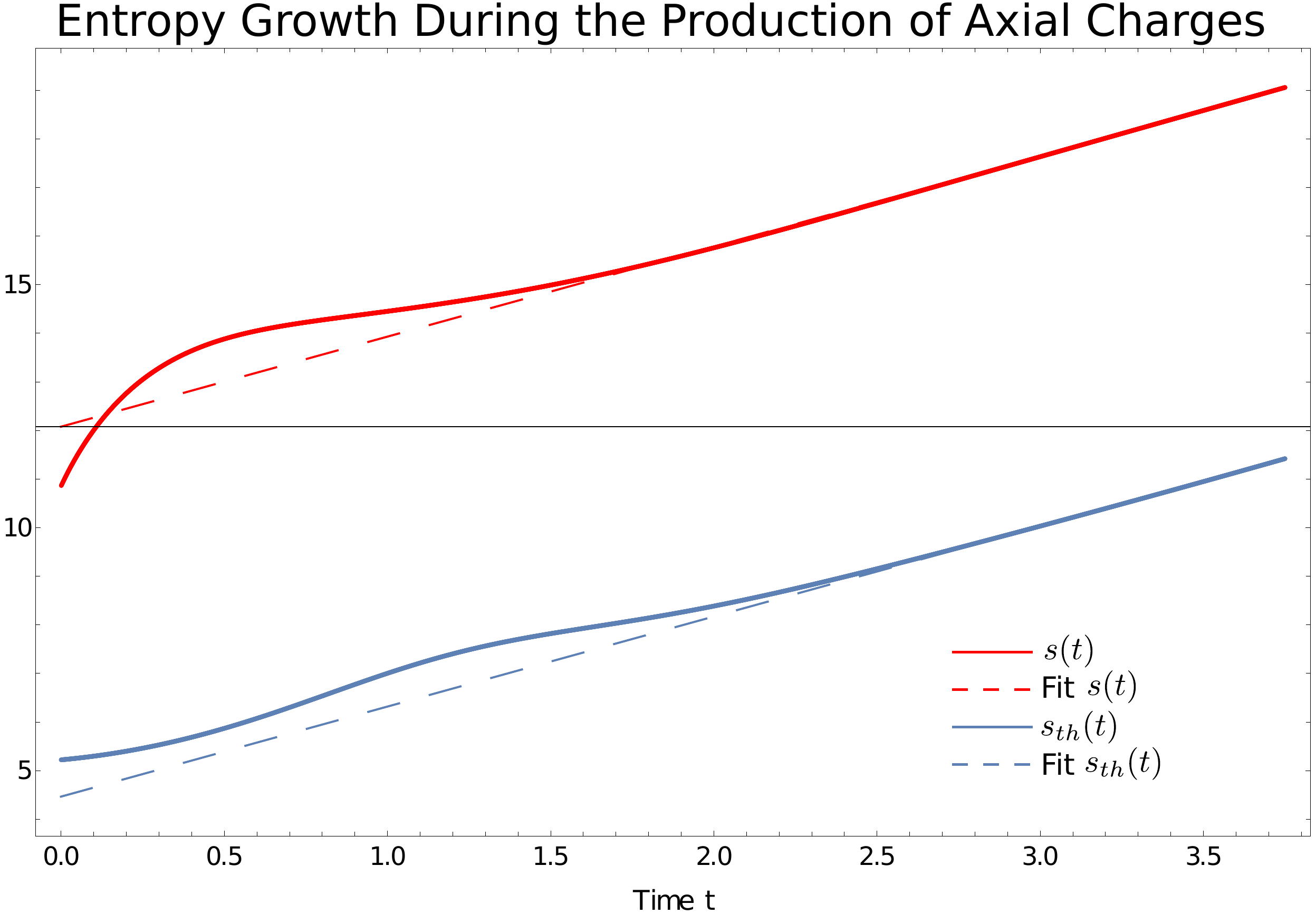}
 \end{center}
\caption{\label{fig:Entropy_Comparision}The thermodynamic entropy as calculated via eq.\ (\ref{eqn:Thermodynamic}) displayed along with the entropy as calculated as the area of the apparent horizon via eq.\ (\ref{eqn:Apparent_Entropy}).
}
\end{figure}

We display in figure~\ref{fig:Entropy_Comparision} the result of matching the growth rate of the thermodynamic entropy to the growth rate of the horizon as measured by the apparent horizon area. The value of the coefficient required for this matching is $a=0.437209$. We can now provide an interpretation of the behavior of both of these measures of entropy. The linear growth can be attributed to the energy delivered to the system via the Joule heating of the plasma.

\section{Summary and Discussion}
In this work we compute for the first time the dynamical evolution of a charged strongly coupled far from equilibrium plasma with a chiral anomaly subjected to external electromagnetic fields. We have computed this evolution as a numerical solution to the Einstein-Maxwell-Chern-Simons equations for an asymptotically anti-de-Sitter spacetime in five dimensions. We have (for the first time in asymptotically AdS$_5$ spacetimes to the author's knowledge) included the dynamical equations for the gauge field into the characteristic formulation of the Einstein equations and evolved them in time alongside the metric components (see eq.\ (\ref{eqn:AS}) to eq.\ (\ref{eqn:ASdotdot})). We have computed the one point functions of the field theory energy-momentum tensor dual to the evolving metric and axial current dual to the evolving bulk gauge field (see eq.\ (\ref{eqn:enmomE}) to eq.\ (\ref{eqn:enmomPl}) and eq.\ (\ref{eqn:Current})). We have displayed the axial current density in the simplest dynamical setup possible to capture the evolution of the current generated due to the axial anomaly during the isotropization of a plasma. Our setup was chosen to mimic conditions found in heavy ion collisions. We have found that aligning external electric and magnetic fields in a plasma with an axial anomaly leads to an azimuthally symmetric inflow of axial charge towards the $x_3$-axis. (see figure~\ref{fig:CurrentEvolution}). This current inflowing from infinity can be considered the source of the generated axial charges which are accelerated along the direction of the electric field (see figure~\ref{fig:Enmom}). As the system evolves the current aligns itself as a flow along the $x_3$-axis beginning along the $x_3$-axis itself and moving azimuthally outward along the cylindrical coordinate $r_c=\sqrt{x_1^2+x_2^2}$ due to the electric field.

At the onset of this work we had hoped to reveal new dynamics associated with a thermalizing charged strongly coupled plasma subjected to electromagnetic fields. The behavior of the current $\braket{J^3}$ in some respects has fulfilled this request. It was not possible to extract this behavior without the full evolution of the system. However we can learn from figure~\ref{fig:gamma_0} that early time evolution of the dual current with and without an axial anomaly displays essentially the same behavior. Therefore our simple model suggests early time behavior of observables (including energy, pressure and current density) of out of equilibrium plasmas with an axial anomaly does not deviate significantly from the evolution without an axial anomaly. It can be see in figure~\ref{fig:gamma_0} the effect of the axial anomaly takes time to produce pronounced deviations from its anomaly free counterpart. 
Furthermore another take away from this work is once again the power of hydrodynamics. Two of our dynamical equations reduce exactly to hydrodynamic predictions. This is perhaps not surprising considering the contribution of the anomaly is exact. In fact the differences in our model with and without a chiral anomaly can be traced back to eq.\ (\ref{eq:energy})-(\ref{eqn:hydroheatcurrent}). Where we see a determining factor in the size of the rate of change of both the energy density and the heat current depend on the size of the current density. With no anomaly present eq.\ (\ref{eq:charge}) enforces the system to have a constant charge density which leads to a constant growth of the heat current as can be see in figure~\ref{fig:gamma_0}. The constant value of the charge density in turn reduces the growth of the energy and in the late time regime this growth appears linear suggesting a stable equilibrium current is being established.

While the early time dynamics is still out of reach of hydrodynamic predictions, hydrodynamics was able to predict nearly all of our systems late time behavior. In spite of this our work represents a necessary step towards the full dynamic evolution of bulk Einstein-Maxwell-Chern-Simons theory needed to provide a meaningful model of the early time dynamics of heavy ion collisions. 

As an application of our solutions we have computed the evolution of the entropy and entanglement entropy. We have found the production and acceleration of axial charges by the electric field leads to the linear growth of both entropy and entanglement entropy in the late time (see figure~\ref{fig:Entropy_With_Axial}). We have also shown that by utilizing a common measure of out of equilibrium temperature the thermodynamic entropy as defined in eq.\ (\ref{eqn:Thermodynamic}) displays exactly the same late time behavior as the horizon entropy. As discussed in section~\ref{sec:Results} we interpret this as further evidence that the late time linear growth in entropy is due to Joule heating during the production of axial charges. We also provided linear fits of the entanglement entropy which encode a notion of entanglement velocity. Turning off the Chern-Simons coupling leads to an increased rate of growth of the thermal entropy. While the thermal entropy growth increases as we turn off the Chern-Simons coupling the rate of growth of the entanglement entropy decreases (see figure~\ref{fig:Entropy_comparison} and table~\ref{tab:FitParameters}). The minimal surfaces we calculate eventually cross the apparent horizon (see figure~\ref{fig:surfaces_Length_dep}). Bulk minimal surfaces used to calculate entanglement entropy passing the apparent horizon were first displayed in a AdS$_3$ Vaidya setup~\cite{AbajoArrastia:2010yt}. And more recently again in AdS$_3$ in the context of the quantum null energy condition~\cite{Ecker:2019ocp}\footnote{We thank the referee for pointing us to this recent work.}. It would be interesting to study further the minimal surfaces and associated entanglement entropy calculated in this work. 


Looking to the future there are many interesting avenues we can now explore however we will mention just three possible directions:

We are interested in finding a simple holographic model in which we can further study the production of axial charges and the CME in an analytic setting. Luckily there have been many works targeting the holographic Schwinger effect~\cite{Semenoff:2011ng,Gorsky:2001up,Ambjorn:2011wz,Kawai:2013xya,Yee:2009vw}. In~\cite{Sato:2013pxa} the authors consider extending this calculation for the inclusion of magnetic fields both perpendicular and parallel to the electric fields. Continuing their work to study the holographic entanglement entropy in a theory producing axial charges in this setting would be a logical continuation of this work. 

Our discussion of the generation of axial currents naturally leads us to the topic of chiral transport. There have been many works interested in chiral transport phenomena (some excellent examples~\cite{Lin:2013sga,Ammon:2016fru,Haack:2018ztx,Pendas:2019}). The author is current engaged in studying these effects far from equilibrium in anisotropic systems. 

In the current work we have static electric and magnetic fields. This is not the case in heavy ion collisions where the electromagnetic fields generated during collisions are highly time dependent~\cite{Skokov:2009qp}. Recent works have tried to address the effect of time dependent electromagnetic fields on heavy ion collisions~\cite{Bali:2011qj,Zhong:2014cda,Pang:2016yuh,Gursoy:2018yai,Ye:2018jwq}. It would be very interesting to extend our current work to include time dependent electromagnetic fields. The Bianchi identity in the Maxwell sector is no longer trivially satisfied when we include time-dependent magnetic fields. This leads to a significantly more complex evolution. However if we want to provide a meaningful comparison to heavy ion collisions this is a necessary step. It will also be necessary to include both a vector and axial gauge field rather then just the axial gauge field displayed in this work. Furthermore as seen in eq.\ (\ref{eqn:Current}) the charge density in the system grows without bound. As discussed in section~\ref{sec:Asymptotic} this behavior is in part an artifact of the translational invariance of our system. Our homogeneous system is essentially a perfect conductor. To alleviate this issue we could introduce momentum relaxation via massive gravity models~\cite{Vegh:2013sk,Blake:2013owa}, Q-Lattice models~\cite{Donos:2014cya} or through the addition of a mass term to gauge field along with use of the St\"uckelberg mechanism~\cite{Jimenez-Alba:2014iia,Klebanov:2002gr}. In addition to the introduction of finite resistivity of the plasma or realistic time dependence of the gauge fields, the gauge fields should also be dynamic rather then external fields. Recent work has displayed it is possible to include fully dynamic gauge fields in the dual field theory picture~\cite{Grozdanov:2016tdf,Grozdanov:2016zjj,Grozdanov:2017kyl,Grozdanov:2018fic,Armas:2018atq,Armas:2018zbe} allowing us to compute, in principle, gauge field correlation functions.\footnote{The author is currently engaged in studying the various extensions mentioned in this paragraph.}

\acknowledgments
The author wishes to thank and acknowledge the following individuals. Matthias Kaminski for technical discussions, comments on an early draft of this work and the encouragement to complete this project independently. Thank you Matthias. Larry Yaffe for discussions about this work during the Holographic QCD (HQCD) conference at NORDITA in Stockholm Sweden. The organizers of HQCD for an engaging conference and lively discussions. Dirk Rischke and Goethe University for the opportunity to present previous work and for local accommodations in Frankfurt am Main where part of this current work was completed. Dmitri Kharzeev for a brief discussion about this work and his suggestion that entanglement entropy would be interesting to study in this system. Sa\v so Grozdanov for helpful comments on a draft of this work. Karl Landsteiner for helpful comments on a draft of this work. The author would also like to thank the referee, his/her comments were very useful and have greatly improved this work. This work was partially supported by the U. S. Department of Energy grant DE-SC-0012447.

\appendix
\section{Scaling Relations}\label{sec:scaling}
It is useful to consider independent scalings of field theory directions spanned by $\mathbf{x}$ and r given by,
\begin{equation}
  \mathbf{x}=\alpha  \tilde{\mathbf{x}}\qquad r=\alpha^{-1}\psi^2\tilde{r}.\label{eqn:scale}
\end{equation}
These rescalings were used in~\cite{Fuini:2015hba} to demonstrate the independence of the field theory from the AdS radius $L$ without the presence of a Chern-Simons term. Due to the omission of this term it is worth our time to verify that with this additional boundary term we again find our field theory to be independent of changes in $L$.

Inspection of line element reveals the scalings eq.\ (\ref{eqn:scale}) will produce an overall conformal factor of the line element if the metric components transform as,
\begin{align}
    \tilde{B}(\tilde{\mbf{x}},\tilde{r})&=B(\mbf{x}(\tilde{\mbf{x}}),r(\tilde{r})),\\
     \tilde{S}(\tilde{\mbf{x}},\tilde{r})&=\frac{\alpha}{\psi}S(\mbf{x}(\tilde{\mbf{x}}),r(\tilde{r})),\\
     \tilde{A}(\tilde{\mbf{x}},\tilde{r})&=\frac{\alpha^2}{\psi^2}A(\mbf{x}(\tilde{\mbf{x}}),r(\tilde{r})),\\
     \tilde{F}(\tilde{\mbf{x}},\tilde{r})&=\frac{\alpha^2}{\psi^2}F(\mbf{x}(\tilde{\mbf{x}}),r(\tilde{r})).
\end{align}
Along with the metric tensor components the transformation also effects the gauge field $\mathcal{A}_{\mu}$ whose components transform as,
\begin{align}
  \tilde{\phi}(\tilde{\mbf{x}},\tilde{r})&=\alpha\phi(\mbf{x}(\tilde{\mbf{x}}),r(\tilde{r})),\\
   \tilde{P}(\tilde{\mbf{x}},\tilde{r})&=\alpha P(\mbf{x}(\tilde{\mbf{x}}),r(\tilde{r})).
\end{align}
Finally we must additionally transform the parameters as follows,
\begin{equation}
    \tilde{\rho}=\alpha^3\rho,\quad\tilde{\mB}=\alpha^2\mB,\quad\tilde{L}=\psi^{-1}L,\quad \tilde{\gamma}=\gamma.
\end{equation}
Performing the scaling transformation on the action shows, $\tilde{S}=S$, hence the action is invariant with respect to these scalings.
Clearly our scaling transformation has no effect on the equations of motion. We can therefore independently scale the AdS radius without changing the boundary theory by taking $\alpha=1,\psi\neq 1$ hence justifying our choice of setting $L=1$.

\section{Residual Symmetries}
\label{sec:Appendix_symm}
We can ask in this system if the radial shift $r\rightarrow r+\xi(v)$ is still a diffeomorphism of the system when the Chern-Simons term is included. Performing the transformation in the line element we find,
\begin{align}
 \exd s^2\rightarrow \exd s'^2&=\omega'\exd v  +S(v,r'-\xi)^2\left(e^{B(v,r'-\xi)}\left(\exd x_1^2+\exd 
x_2^2\right)+e^{-2B(v,r'-\xi)}\exd x_3^2\right), \\
\omega'&=(-A(v,r'-\xi)\exd v+F(v,r'-\xi)\exd x_3+2\exd r'-2\xi'(v)\exd v).
\end{align}
The line element will be invariant if $\tilde{A}(v,r')=A(v,r'-\xi)+2\xi'(v)$, exactly as described by the authors in~\cite{Chesler:2013lia}. The form of our gauge field is invariant under the transformation,
\begin{equation}
\mathcal{A}_{\mu}(v,r'-\xi)\exd x'^{\mu}=\phi(v,r'-\xi)\exd t+\frac{1}{2}x_2\mathcal{B}\exd x_1-\frac{1}{2}x_1\mathcal{B}\exd x_2-
P(v,r'-\xi)\exd x_3.
\end{equation}
With both the transformation of the metric components and the gauge field one can show explicitly that the action eq.\ (\ref{CS-System}) is invariant under bulk radial diffeomorphism and hence $r\rightarrow r+\xi$ is still a good symmetry. 

Although the action is invariant under bulk radial shifts looking at eq.\ (\ref{eqn:Current}) it is not immediately evident that the coefficient $p_2(t)$ is independent of the radial gauge transformation. It cannot be the case that the dual current depends on the bulk radial shift as the action is clearly gauge invariant. The issue is the explicit presence of the Eddington-Finkelstein time $v$ in the expressions. This quantity is a function of the radial coordinate $v(r)$, obscuring the gauge dependence. We can write $v=t-r^*\approx t-1/r$, keeping only the leading order contribution as we are interested in near boundary behavior, inserting this expression into eq.\ (\ref{eqn:nbeP}) we see,
\begin{align}
 P(t,r)&=p_0+E\left(t-\frac{1}{r}\right) + \frac{E}{r}+\frac{p_2(t)}{r^2}+\cdots \, , \\
 &=p_0+E t +\frac{p_2(t)}{r^2}+\cdots, \label{eqn:poincareP}
\end{align}
where the dots indicate terms higher order in $1/r$. It is the quantity at order $1/r^2$ in eq.\ (\ref{eqn:poincareP}) which is to be considered dual to $\braket{J^3}$ (asides from contributions from the Chern-Simons action). Performing a radial gauge transformation $r\rightarrow r+\xi$ and again expanding the result near the conformal boundary returns exactly the same expression as that given in eq.\ (\ref{eqn:poincareP}) thus leaving the $x_3$ component of the dual current invariant under bulk radial shifts. In practice this quantity can be extracted by subtracting terms up to order $O(1/r^2)$ and working with a field scaled by $1/r^2$,
\begin{equation}
 P(v,r)=p_0+E\left(v+\frac{1}{r}\right)+\frac{P_s(v,r)}{r^2}.
\end{equation}
The function behaves as $P_s=O(r^0)$ near the conformal boundary, $r\rightarrow\infty$, leaving us with the ability to extract the necessary coefficient $p_2(t)$ via,
\begin{equation}
 p_2(t)=\lim_{r\rightarrow\infty} P_s(v,r),
\end{equation}
where we recall again that $v({r\rightarrow\infty})=t$.

\section{Equations of Motion}
\label{sec:Appendix_eqns}
The equations of motion that result from the action can be written as a partially nested list. As described in section~\ref{sec:Numerical_Tech} we can solve the first three equations in turn. These equations are,
\begin{align}
    6 S S''&=-3 S^2 \left(B'\right)^2-4 e^{2 B}
  \left(P'\right)^2 \label{eqn:AS}\\
3 S^3 F ''&=-3 S^2 S' \left(6 F 
  B'+F '\right)-3 S^3 \left(2 B' F
  '+F  \left(2
  B''+\left(B'\right)^2\right)\right)\nonumber \\
  &-4 S
  F  \left(e^{2 B} \left(P'\right)^2-3
  \left(S'\right)^2\right)+12 P' (\gamma  P
  \mathcal{B}+\rho ) \label{eqn:AF}\\
 12 e^{2 B} S^5 \dot{S}'&=-e^{2 B} S^4
  \left(e^{2 B} F ^2 \left(B'\right)^2+4
  e^{2 B} F  B' F '+e^{2 B} \left(F
  '\right)^2+24 \dot{S} S'\right)\nonumber \\
  &-4 e^{4 B}
  S^3 F  S' \left(F  B'+2 F
  '\right)-4 e^{2 B} \gamma ^2 P^2
  \mathcal{B}^2 \nonumber\\&-4 S^2 \left(F ^2
  \left(e^{4 B} \left(S'\right)^2-e^{6 B}
  \left(P'\right)^2\right)+\mathcal{B}^2\right)
 \nonumber \\ &-8 e^{2 B} \gamma  P \rho 
  \mathcal{B}-4 e^{2 B} \rho ^2+24 e^{2 B}
  S^6 \label{eqn:ASdot}
  \end{align}
  The next set is a coupled system for $\dot{P}$ and $\dot{B}$.
  \begin{align}
 2 e^{2 B} S^4 \dot{P}'&=-2 e^{2 B} S^4
  \left(\dot{P} B'+\dot{B} P'\right)+e^{2 B}
  S \left(2 \rho  F  B'+e^{2 B} F ^2
  P' S'+\rho  F '\right)\nonumber \\
  &-e^{4 B} S^2 F
   \left(F  \left(4 B' P'+P''\right)+2 P'
  F '\right)\nonumber \\
  &+\gamma  P \mathcal{B}
  \left(e^{2 B} S \left(2 F  B'+F
  '\right)-2 e^{2 B} F  S'+\gamma 
  \mathcal{B}\right)\nonumber \\
  &-e^{2 B} S^3
  \left(\dot{S} P'+\dot{P} S'\right)+\rho 
  \left(\gamma  \mathcal{B}-2 e^{2 B} F 
  S'\right) \label{eqn:APdot}\\
 6 e^{2 B} S^4 \dot{B}'&=-e^{4 B} S F  S'
  \left(11 F  B'+4 F '\right)-9 e^{2
  B} S^3 \left(\dot{S} B'+\dot{B}
  S'\right)\nonumber \\
  &-e^{4 B} S^2 \left(2 F  B'
  F '+F ^2 \left(3 B''+2
  \left(B'\right)^2\right)-8 \dot{P}
  P'-\left(F '\right)^2\right)\nonumber\\
  &+4 e^{4 B}
  F ^2 \left(e^{2 B}\left(P'\right)^2+\left(S'\right)^2\right)  -4 \mathcal{B}^2 \label{eqn:ABdot}
  \end{align}
 The next two equations for $A$ and $\dot{F}$ can be solved as a nested list again. 
\begin{align}
 6 e^{2 B} S^6 A''&=-6 e^{2 B} S^6 \left(3\dot{B} B'+4\right)+24 e^{4 B} S^3 F S' \left(2 F  B'+F '\right)\nonumber \\
  &+e^{2 B}S^4 \left(12 e^{2 B} F  B' F '+3e^{2 B} F ^2 \left(4 B''+3\left(B'\right)^2\right)\right)\nonumber \\
 &+e^{2 B} S^4  \left( -8 e^{2 B} \dot{P}
  P'-3 e^{2 B} \left(F '\right)^2+72
  \dot{S} S'\right)+28 e^{2 B} \gamma ^2 P^2
  \mathcal{B}^2 \nonumber \\
  &+4 S^2 \left(e^{4 B} F ^2
  \left(e^{2 B} \left(P'\right)^2-3
  \left(S'\right)^2\right)+5
  \mathcal{B}^2\right)-48 e^{4 B} \rho  S
  F  P' \nonumber \\
  &-8 e^{2 B} \gamma  P \mathcal{B}
  \left(6 e^{2 B} S F  P'-7 \rho
  \right)+28 e^{2 B} \rho ^2 \label{eqn:AA}
\end{align}
\begin{align}
 6 e^{2 B} S^6 \dot{F }'&=-F  \left(-6
   e^{2 B} S^6 \left(A' B'+4\right)+6 e^{2 B}
  S^5 \left(A' S'+3 \dot{S} B'-3 \dot{B}
  S'\right)\nonumber \right. \\
  &\left.+4 e^{2 B} \gamma ^2 P^2
  \mathcal{B}^2+e^{4 B} S^4 \left(16 \dot{P}
  P'+3 \left(F '\right)^2\right)+8 e^{2
  B} \gamma  P \rho  \mathcal{B}+4 e^{2 B}
  \rho ^2-4 S^2 \mathcal{B}^2\right)\nonumber\\
  &+3 e^{2
  B} S^3 \left(S^3 \left(A' F '-4
  \dot{F } B'\right)+8 \dot{P} (\gamma  P
  \mathcal{B}+\rho )+S^2 \left(4 \dot{F }
  S'-6 \dot{S} F '\right)\right)\nonumber \\
  &-e^{4 B}
   S^2 F ^3 \left(-18 S B' S'-3 S^2
  \left(2 B''+\left(B'\right)^2\right)+4
  \left(e^{2 B} \left(P'\right)^2+3
  \left(S'\right)^2\right)\right) \label{eqn:AFdot}
  \end{align}
  The final equation is a constraint equation which can be used to monitor the accuracy of the code. 
  \begin{align}
 12 S^7 \ddot{S}&=F ^2 \left(6 e^{2 B} S^5
  \left(A' S'+2 \dot{S} B'-2 \dot{B}
  S'\right)-6 e^{2 B} S^6 \left(\dot{B}
  B'+4\right)+4 e^{2 B} \gamma ^2 P^2
  \mathcal{B}^2\nonumber \right.\\
  &\left.+e^{2 B} S^4 \left(-8 e^{2 B}
  \dot{P} P'+e^{2 B} \left(F
   '\right)^2+24 \dot{S} S'\right)+8 e^{2 B}
  \gamma  P \rho  \mathcal{B}+4 e^{2 B} \rho
  ^2+4 S^2 \mathcal{B}^2\right)\nonumber \\
  &-2 S^6
  \left(-3 S \dot{S} A'+4 e^{2 B}
  \dot{P}^2+3 \dot{B}^2 S^2\right)\nonumber \\
  &-e^{4 B}
  S^2 F ^4 \left(-S^2 \left(B'\right)^2-4
  S B' S'+4 e^{2 B} \left(P'\right)^2-4
  \left(S'\right)^2\right)\nonumber \\
  &+4 e^{4 B} S^3
  F ^3 F ' \left(S B'+2 S'\right)+12
  e^{2 B} S^5 F  \left(\dot{S} F
  '-\dot{F } S'\right) \label{eqn:ASdotdot}
\end{align}
We can confirm that the maximal violation of the constraint was below $10^{-6}$ throughout the entire evolution of the system. The violation of the constraint on average on each time step was on the order $10^{-15}$ throughout the evolution of the system.

\bibliographystyle{JHEP}
\bibliography{iso_pub}

\providecommand{\href}[2]{#2}\begingroup\raggedright\begin{thebibliography}{100}

\bibitem{Kharzeev:2007jp}
D.~E. Kharzeev, L.~D. McLerran and H.~J. Warringa, \emph{The effects of
  topological charge change in heavy ion collisions: 'event by event p and cp
  violation'},
  \href{https://doi.org/10.1016/j.nuclphysa.2008.02.298}{\emph{Nucl. Phys.}
  {\bfseries A803} (2008) 227}
  [\href{https://arxiv.org/abs/0711.0950}{{\ttfamily 0711.0950}}].

\bibitem{Skokov:2009qp}
V.~Skokov, A.~{\relax Yu}. Illarionov and V.~Toneev, \emph{Estimate of the
  magnetic field strength in heavy-ion collisions},
  \href{https://doi.org/10.1142/S0217751X09047570}{\emph{Int. J. Mod. Phys.}
  {\bfseries A24} (2009) 5925}
  [\href{https://arxiv.org/abs/0907.1396}{{\ttfamily 0907.1396}}].

\bibitem{Kharzeev:2004ey}
D.~Kharzeev, \emph{Parity violation in hot qcd: Why it can happen, and how to
  look for it},
  \href{https://doi.org/10.1016/j.physletb.2005.11.075}{\emph{Phys. Lett.}
  {\bfseries B633} (2006) 260}
  [\href{https://arxiv.org/abs/hep-ph/0406125}{{\ttfamily hep-ph/0406125}}].

\bibitem{Vilenkin1978}
A.~Vilenkin, \emph{Parity nonconservation and rotating black holes},
  \href{https://doi.org/10.1103/PhysRevLett.41.1575}{\emph{Phys. Rev. Lett.}
  {\bfseries 41} (1978) 1575}.

\bibitem{Warringa:2012bq}
H.~J. Warringa, \emph{Dynamics of the chiral magnetic effect in a weak magnetic
  field}, \href{https://doi.org/10.1103/PhysRevD.86.085029}{\emph{Phys. Rev.}
  {\bfseries D86} (2012) 085029}
  [\href{https://arxiv.org/abs/1205.5679}{{\ttfamily 1205.5679}}].

\bibitem{Voloshin2004}
S.~A. Voloshin, \emph{Parity violation in hot qcd: How to detect it},
  \href{https://doi.org/10.1103/PhysRevC.70.057901}{\emph{Phys. Rev. C}
  {\bfseries 70} (2004) 057901}.

\bibitem{Tang:2019pbl}
A.~H. Tang, \emph{Probe chiral magnetic effect with signed balance function},
  \href{https://arxiv.org/abs/1903.04622}{{\ttfamily 1903.04622}}.

\bibitem{Lin:2020jcp}
{\scshape STAR} collaboration, Y.~Lin, \emph{Measurement of the charge
  separation along the magnetic field with signed balance function in 200 gev
  au + au collisions at star},  2020,
  \href{https://arxiv.org/abs/2002.11446}{{\ttfamily 2002.11446}}.

\bibitem{Kharzeev2019}
D.~E. Kharzeev and J.~Liao, \emph{Isobar collisions at rhic to test local
  parity violation in strong interactions},
  \href{https://doi.org/10.1080/10619127.2018.1495479}{\emph{Nuclear Physics
  News} {\bfseries 29} (2019) 26}.

\bibitem{STAR2009CME}
{\scshape STAR Collaboration} collaboration, B.~I. Abelev, M.~M. Aggarwal,
  Z.~Ahammed, A.~V. Alakhverdyants, B.~D. Anderson, D.~Arkhipkin et~al.,
  \emph{Azimuthal charged-particle correlations and possible local strong
  parity violation},
  \href{https://doi.org/10.1103/PhysRevLett.103.251601}{\emph{Phys. Rev. Lett.}
  {\bfseries 103} (2009) 251601}.

\bibitem{Deng:2016wt}
W.-T. Deng, X.-G. Huang, G.-L. Ma and G.~Wang, \emph{Testing the chiral
  magnetic effect with isobaric collisions},
  \href{https://doi.org/10.1103/PhysRevC.94.041901}{\emph{Phys. Rev. C}
  {\bfseries 94} (2016) 041901}.

\bibitem{Skokov:2016yrj}
V.~Koch, S.~Schlichting, V.~Skokov, P.~Sorensen, J.~Thomas, S.~Voloshin et~al.,
  \emph{Status of the chiral magnetic effect and collisions of isobars},
  \href{https://doi.org/10.1088/1674-1137/41/7/072001}{\emph{Chin. Phys.}
  {\bfseries C41} (2017) 072001}
  [\href{https://arxiv.org/abs/1608.00982}{{\ttfamily 1608.00982}}].

\bibitem{Deng2016}
W.-T. Deng, X.-G. Huang, G.-L. Ma and G.~Wang, \emph{Testing the chiral
  magnetic effect with isobaric collisions},
  \href{https://doi.org/10.1103/PhysRevC.94.041901}{\emph{Phys. Rev. C}
  {\bfseries 94} (2016) 041901}.

\bibitem{Xiong413}
J.~Xiong, S.~K. Kushwaha, T.~Liang, J.~W. Krizan, M.~Hirschberger, W.~Wang
  et~al., \emph{Evidence for the chiral anomaly in the dirac semimetal na3bi},
  \href{https://doi.org/10.1126/science.aac6089}{\emph{Science} {\bfseries 350}
  (2015) 413}
  [\href{https://arxiv.org/abs/https://science.sciencemag.org/content/350/6259/413.full.pdf}{{\ttfamily
  https://science.sciencemag.org/content/350/6259/413.full.pdf}}].

\bibitem{AbajoArrastia:2010yt}
J.~Abajo-Arrastia, J.~Aparicio and E.~Lopez, \emph{Holographic evolution of
  entanglement entropy},
  \href{https://doi.org/10.1007/JHEP11(2010)149}{\emph{JHEP} {\bfseries 11}
  (2010) 149} [\href{https://arxiv.org/abs/1006.4090}{{\ttfamily 1006.4090}}].

\bibitem{Balasubramanian:2011ur}
V.~Balasubramanian, A.~Bernamonti, J.~de~Boer, N.~Copland, B.~Craps,
  E.~Keski-Vakkuri et~al., \emph{Holographic thermalization},
  \href{https://doi.org/10.1103/PhysRevD.84.026010}{\emph{Phys. Rev.}
  {\bfseries D84} (2011) 026010}
  [\href{https://arxiv.org/abs/1103.2683}{{\ttfamily 1103.2683}}].

\bibitem{Caceres:2012em}
E.~Caceres and A.~Kundu, \emph{Holographic thermalization with chemical
  potential}, \href{https://doi.org/10.1007/JHEP09(2012)055}{\emph{JHEP}
  {\bfseries 09} (2012) 055} [\href{https://arxiv.org/abs/1205.2354}{{\ttfamily
  1205.2354}}].

\bibitem{Ebrahim:2010ra}
H.~Ebrahim and M.~Headrick, \emph{Instantaneous thermalization in holographic
  plasmas},  \href{https://arxiv.org/abs/1010.5443}{{\ttfamily 1010.5443}}.

\bibitem{Keranen:2011xs}
V.~Keranen, E.~Keski-Vakkuri and L.~Thorlacius, \emph{Thermalization and
  entanglement following a non-relativistic holographic quench},
  \href{https://doi.org/10.1103/PhysRevD.85.026005}{\emph{Phys. Rev.}
  {\bfseries D85} (2012) 026005}
  [\href{https://arxiv.org/abs/1110.5035}{{\ttfamily 1110.5035}}].

\bibitem{Camilo:2014npa}
G.~Camilo, B.~Cuadros-Melgar and E.~Abdalla, \emph{Holographic thermalization
  with a chemical potential from born-infeld electrodynamics}, {\emph{JHEP}
  {\bfseries 02} (2015) 103}.

\bibitem{Hu:2016mym}
Y.-P. Hu, X.-X. Zeng and H.-Q. Zhang, \emph{Holographic thermalization and
  generalized vaidya-ads solutions in massive gravity},
  \href{https://doi.org/10.1016/j.physletb.2016.12.028}{\emph{Phys. Lett.}
  {\bfseries B765} (2017) 120}
  [\href{https://arxiv.org/abs/1611.00677}{{\ttfamily 1611.00677}}].

\bibitem{Giordano:2014kya}
A.~Giordano, N.~E. Grandi and G.~A. Silva, \emph{Holographic thermalization of
  charged operators},
  \href{https://doi.org/10.1007/JHEP05(2015)016}{\emph{JHEP} {\bfseries 05}
  (2015) 016} [\href{https://arxiv.org/abs/1412.7953}{{\ttfamily 1412.7953}}].

\bibitem{Zhang:2015dia}
S.-J. Zhang and E.~Abdalla, \emph{Holographic thermalization in charged dilaton
  anti-de sitter spacetime},
  \href{https://doi.org/10.1016/j.nuclphysb.2015.05.005}{\emph{Nucl. Phys.}
  {\bfseries B896} (2015) 569}
  [\href{https://arxiv.org/abs/1503.07700}{{\ttfamily 1503.07700}}].

\bibitem{Galante:2012pv}
D.~Galante and M.~Schvellinger, \emph{Thermalization with a chemical potential
  from ads spaces}, \href{https://doi.org/10.1007/JHEP07(2012)096}{\emph{JHEP}
  {\bfseries 07} (2012) 096} [\href{https://arxiv.org/abs/1205.1548}{{\ttfamily
  1205.1548}}].

\bibitem{Dey:2015poa}
A.~Dey, S.~Mahapatra and T.~Sarkar, \emph{Holographic thermalization with weyl
  corrections}, \href{https://doi.org/10.1007/JHEP01(2016)088}{\emph{JHEP}
  {\bfseries 01} (2016) 088}
  [\href{https://arxiv.org/abs/1510.00232}{{\ttfamily 1510.00232}}].

\bibitem{Arefeva:2012jp}
I.~{\relax Ya}. Arefeva and I.~V. Volovich, \emph{On holographic thermalization
  and dethermalization of quark-gluon plasma},
  \href{https://arxiv.org/abs/1211.6041}{{\ttfamily 1211.6041}}.

\bibitem{Atashi:2016fai}
M.~Atashi, K.~Bitaghsir~Fadafan and G.~Jafari, \emph{Linearized holographic
  isotropization at finite coupling},
  \href{https://doi.org/10.1140/epjc/s10052-017-4995-2}{\emph{Eur. Phys. J.}
  {\bfseries C77} (2017) 430}
  [\href{https://arxiv.org/abs/1611.09295}{{\ttfamily 1611.09295}}].

\bibitem{Zhang:2014cga}
S.-J. Zhang, B.~Wang, E.~Abdalla and E.~Papantonopoulos, \emph{Holographic
  thermalization in gauss-bonnet gravity with de sitter boundary},
  \href{https://doi.org/10.1103/PhysRevD.91.106010}{\emph{Phys. Rev.}
  {\bfseries D91} (2015) 106010}
  [\href{https://arxiv.org/abs/1412.7073}{{\ttfamily 1412.7073}}].

\bibitem{Ageev:2017wet}
D.~S. Ageev and I.~{\relax Ya}. Aref'eva, \emph{Holographic non-equilibrium
  heating}, \href{https://doi.org/10.1007/JHEP03(2018)103}{\emph{JHEP}
  {\bfseries 03} (2018) 103}
  [\href{https://arxiv.org/abs/1704.07747}{{\ttfamily 1704.07747}}].

\bibitem{Andrade:2016rln}
T.~Andrade, J.~Casalderrey-Solana and A.~Ficnar, \emph{Holographic
  isotropisation in gauss-bonnet gravity},
  \href{https://doi.org/10.1007/JHEP02(2017)016}{\emph{JHEP} {\bfseries 02}
  (2017) 016} [\href{https://arxiv.org/abs/1610.08987}{{\ttfamily
  1610.08987}}].

\bibitem{Wondrak:2017kgp}
M.~F. Wondrak, M.~Kaminski, P.~Nicolini and M.~Bleicher, \emph{{AdS/CFT far
  from equilibrium in a Vaidya setup}},
  \href{https://doi.org/10.1088/1742-6596/942/1/012020}{\emph{J. Phys. Conf.
  Ser.} {\bfseries 942} (2017) 012020}
  [\href{https://arxiv.org/abs/1711.08835}{{\ttfamily 1711.08835}}].

\bibitem{Wondrak:2020tzt}
M.~F. Wondrak, M.~Kaminski and M.~Bleicher, \emph{Shear transport far from
  equilibrium via holography},
  \href{https://arxiv.org/abs/2002.11730}{{\ttfamily 2002.11730}}.

\bibitem{Chesler:2008hg}
P.~M. Chesler and L.~G. Yaffe, \emph{Horizon formation and far-from-equilibrium
  isotropization in supersymmetric yang-mills plasma},
  \href{https://doi.org/10.1103/PhysRevLett.102.211601}{\emph{Phys. Rev. Lett.}
  {\bfseries 102} (2009) 211601}
  [\href{https://arxiv.org/abs/0812.2053}{{\ttfamily 0812.2053}}].

\bibitem{Chesler:2010bi}
P.~M. Chesler and L.~G. Yaffe, \emph{Holography and colliding gravitational
  shock waves in asymptotically {$AdS_5$} spacetime},
  \href{https://doi.org/10.1103/PhysRevLett.106.021601}{\emph{Phys. Rev. Lett.}
  {\bfseries 106} (2011) 021601}
  [\href{https://arxiv.org/abs/1011.3562}{{\ttfamily 1011.3562}}].

\bibitem{vanderSchee:2012qj}
W.~van~der Schee, \emph{Holographic thermalization with radial flow},
  \href{https://doi.org/10.1103/PhysRevD.87.061901}{\emph{Phys. Rev.}
  {\bfseries D87} (2013) 061901}
  [\href{https://arxiv.org/abs/1211.2218}{{\ttfamily 1211.2218}}].

\bibitem{vanderSchee:2013pia}
W.~van~der Schee, P.~Romatschke and S.~Pratt, \emph{Fully dynamical simulation
  of central nuclear collisions},
  \href{https://doi.org/10.1103/PhysRevLett.111.222302}{\emph{Phys. Rev. Lett.}
  {\bfseries 111} (2013) 222302}
  [\href{https://arxiv.org/abs/1307.2539}{{\ttfamily 1307.2539}}].

\bibitem{Casalderrey-Solana:2013aba}
J.~Casalderrey-Solana, M.~P. Heller, D.~Mateos and W.~van~der Schee, \emph{From
  full stopping to transparency in a holographic model of heavy ion
  collisions},
  \href{https://doi.org/10.1103/PhysRevLett.111.181601}{\emph{Phys. Rev. Lett.}
  {\bfseries 111} (2013) 181601}
  [\href{https://arxiv.org/abs/1305.4919}{{\ttfamily 1305.4919}}].

\bibitem{Casalderrey-Solana:2016xfq}
J.~Casalderrey-Solana, D.~Mateos, W.~van~der Schee and M.~Triana,
  \emph{Holographic heavy ion collisions with baryon charge},
  \href{https://doi.org/10.1007/JHEP09(2016)108}{\emph{JHEP} {\bfseries 09}
  (2016) 108} [\href{https://arxiv.org/abs/1607.05273}{{\ttfamily
  1607.05273}}].

\bibitem{Chesler:2015wra}
P.~M. Chesler and L.~G. Yaffe, \emph{Holography and off-center collisions of
  localized shock waves},
  \href{https://doi.org/10.1007/JHEP10(2015)070}{\emph{JHEP} {\bfseries 10}
  (2015) 070} [\href{https://arxiv.org/abs/1501.04644}{{\ttfamily
  1501.04644}}].

\bibitem{Grozdanov:2016zjj}
S.~Grozdanov and W.~van~der Schee, \emph{Coupling constant corrections in a
  holographic model of heavy ion collisions},
  \href{https://doi.org/10.1103/PhysRevLett.119.011601}{\emph{Phys. Rev. Lett.}
  {\bfseries 119} (2017) 011601}
  [\href{https://arxiv.org/abs/1610.08976}{{\ttfamily 1610.08976}}].

\bibitem{Waeber:2019nqd}
S.~Waeber, A.~Rabenstein, A.~Schäfer and L.~G. Yaffe, \emph{Asymmetric
  shockwave collisions in ads$_5$},
  \href{https://doi.org/10.1007/JHEP08(2019)005}{\emph{JHEP} {\bfseries 08}
  (2019) 005} [\href{https://arxiv.org/abs/1906.05086}{{\ttfamily
  1906.05086}}].

\bibitem{Muller:2020ziz}
B.~Müller, A.~Rabenstein, A.~Schäfer, S.~Waeber and L.~G. Yaffe,
  \emph{Phenomenological implications of asymmetric $ads_5$ shockwave collision
  studies for heavy ion physics},
  \href{https://arxiv.org/abs/2001.07161}{{\ttfamily 2001.07161}}.

\bibitem{D'Hoker:2009mm}
E.~D'Hoker and P.~Kraus, \emph{Magnetic brane solutions in ads},
  \href{https://doi.org/10.1088/1126-6708/2009/10/088}{\emph{JHEP} {\bfseries
  0910} (2009) 088} [\href{https://arxiv.org/abs/0908.3875}{{\ttfamily
  0908.3875}}].

\bibitem{DHoker:2009ixq}
E.~D'Hoker and P.~Kraus, \emph{Charged magnetic brane solutions in ads (5) and
  the fate of the third law of thermodynamics},
  \href{https://doi.org/10.1007/JHEP03(2010)095}{\emph{JHEP} {\bfseries 03}
  (2010) 095} [\href{https://arxiv.org/abs/0911.4518}{{\ttfamily 0911.4518}}].

\bibitem{DHoker:2010onp}
E.~D'Hoker and P.~Kraus, \emph{Magnetic field induced quantum criticality via
  new asymptotically ads$_5$ solutions},
  \href{https://doi.org/10.1088/0264-9381/27/21/215022}{\emph{Class. Quant.
  Grav.} {\bfseries 27} (2010) 215022}
  [\href{https://arxiv.org/abs/1006.2573}{{\ttfamily 1006.2573}}].

\bibitem{Janiszewski:2015ura}
S.~Janiszewski and M.~Kaminski, \emph{Quasinormal modes of magnetic and
  electric black branes versus far from equilibrium anisotropic fluids},
  \href{https://doi.org/10.1103/PhysRevD.93.025006}{\emph{Phys. Rev.}
  {\bfseries D93} (2016) 025006}
  [\href{https://arxiv.org/abs/1508.06993}{{\ttfamily 1508.06993}}].

\bibitem{Endrodi:2018ikq}
G.~Endrodi, M.~Kaminski, A.~Schafer, J.~Wu and L.~Yaffe, \emph{Universal
  magnetoresponse in qcd and $\mathcal{N}=4$ sym},
  \href{https://doi.org/10.1007/JHEP09(2018)070}{\emph{JHEP} {\bfseries 09}
  (2018) 070} [\href{https://arxiv.org/abs/1806.09632}{{\ttfamily
  1806.09632}}].

\bibitem{Fuini:2015hba}
J.~F. Fuini and L.~G. Yaffe, \emph{Far-from-equilibrium dynamics of a strongly
  coupled non-abelian plasma with non-zero charge density or external magnetic
  field}, \href{https://doi.org/10.1007/JHEP07(2015)116}{\emph{JHEP} {\bfseries
  07} (2015) 116} [\href{https://arxiv.org/abs/1503.07148}{{\ttfamily
  1503.07148}}].

\bibitem{Cartwright2019}
C.~Cartwright and M.~Kaminski, \emph{Correlations far from equilibrium in
  charged strongly coupled fluids subjected to a strong magnetic field},
  \href{https://doi.org/10.1007/JHEP09(2019)072}{\emph{JHEP} {\bfseries 2019}
  (2019) 72}.

\bibitem{Lin:2013sga}
S.~Lin and H.-U. Yee, \emph{Out-of-equilibrium chiral magnetic effect at strong
  coupling}, \href{https://doi.org/10.1103/PhysRevD.88.025030}{\emph{Phys.
  Rev.} {\bfseries D88} (2013) 025030}
  [\href{https://arxiv.org/abs/1305.3949}{{\ttfamily 1305.3949}}].

\bibitem{Pendas:2019}
J.~Fern\'andez-Pend\'as and K.~Landsteiner, \emph{Out-of-equilibrium chiral
  magnetic effect and momentum relaxation in holography},
  \href{https://doi.org/10.1103/PhysRevD.100.126024}{\emph{Phys. Rev. D}
  {\bfseries 100} (2019) 126024}.

\bibitem{Folkestad:2019lam}
A.~Folkestad, S.~Grozdanov, K.~Rajagopal and W.~van~der Schee, \emph{Coupling
  constant corrections in a holographic model of heavy ion collisions with
  nonzero baryon number density},
  \href{https://doi.org/10.1007/JHEP12(2019)093}{\emph{JHEP} {\bfseries 12}
  (2019) 093} [\href{https://arxiv.org/abs/1907.13134}{{\ttfamily
  1907.13134}}].

\bibitem{Gubser:2008pc}
S.~S. Gubser, S.~S. Pufu and A.~Yarom, \emph{Entropy production in collisions
  of gravitational shock waves and of heavy ions},
  \href{https://doi.org/10.1103/PhysRevD.78.066014}{\emph{Phys. Rev.}
  {\bfseries D78} (2008) 066014}
  [\href{https://arxiv.org/abs/0805.1551}{{\ttfamily 0805.1551}}].

\bibitem{Gubser:2009sx}
S.~S. Gubser, S.~S. Pufu and A.~Yarom, \emph{Off-center collisions in ads(5)
  with applications to multiplicity estimates in heavy-ion collisions},
  \href{https://doi.org/10.1088/1126-6708/2009/11/050}{\emph{JHEP} {\bfseries
  11} (2009) 050} [\href{https://arxiv.org/abs/0902.4062}{{\ttfamily
  0902.4062}}].

\bibitem{Lin:2009pn}
S.~Lin and E.~Shuryak, \emph{Grazing collisions of gravitational shock waves
  and entropy production in heavy ion collision},
  \href{https://doi.org/10.1103/PhysRevD.79.124015}{\emph{Phys. Rev.}
  {\bfseries D79} (2009) 124015}
  [\href{https://arxiv.org/abs/0902.1508}{{\ttfamily 0902.1508}}].

\bibitem{Kharzeev:2011ds}
D.~E. Kharzeev and H.-U. Yee, \emph{Anomalies and time reversal invariance in
  relativistic hydrodynamics: the second order and higher dimensional
  formulations}, \href{https://doi.org/10.1103/PhysRevD.84.045025}{\emph{Phys.
  Rev.} {\bfseries D84} (2011) 045025}
  [\href{https://arxiv.org/abs/1105.6360}{{\ttfamily 1105.6360}}].

\bibitem{Ecker:2015kna}
C.~Ecker, D.~Grumiller and S.~A. Stricker, \emph{Evolution of holographic
  entanglement entropy in an anisotropic system},
  \href{https://doi.org/10.1007/JHEP07(2015)146}{\emph{JHEP} {\bfseries 07}
  (2015) 146} [\href{https://arxiv.org/abs/1506.02658}{{\ttfamily
  1506.02658}}].

\bibitem{Ecker:2016thn}
C.~Ecker, D.~Grumiller, P.~Stanzer, S.~A. Stricker and W.~van~der Schee,
  \emph{Exploring nonlocal observables in shock wave collisions},
  \href{https://doi.org/10.1007/JHEP11(2016)054}{\emph{JHEP} {\bfseries 11}
  (2016) 054} [\href{https://arxiv.org/abs/1609.03676}{{\ttfamily
  1609.03676}}].

\bibitem{Bondi:1960jsa}
H.~Bondi, \emph{Gravitational waves in general relativity},
  \href{https://doi.org/10.1038/186535a0}{\emph{Nature} {\bfseries 186} (1960)
  535}.

\bibitem{Sachs:1962wk}
R.~K. Sachs, \emph{Gravitational waves in general relativity. 8. waves in
  asymptotically flat space-times},
  \href{https://doi.org/10.1098/rspa.1962.0206}{\emph{Proc. Roy. Soc. Lond.}
  {\bfseries A270} (1962) 103}.

\bibitem{Ryu:2006bv}
S.~Ryu and T.~Takayanagi, \emph{Holographic derivation of entanglement entropy
  from ads/cft},
  \href{https://doi.org/10.1103/PhysRevLett.96.181602}{\emph{Phys. Rev. Lett.}
  {\bfseries 96} (2006) 181602}
  [\href{https://arxiv.org/abs/hep-th/0603001}{{\ttfamily hep-th/0603001}}].

\bibitem{Lewkowycz:2013nqa}
A.~Lewkowycz and J.~Maldacena, \emph{Generalized gravitational entropy},
  \href{https://doi.org/10.1007/JHEP08(2013)090}{\emph{JHEP} {\bfseries 08}
  (2013) 090} [\href{https://arxiv.org/abs/1304.4926}{{\ttfamily 1304.4926}}].

\bibitem{Konopinski:1978}
E.~J. Konopinski, \emph{What the electromagnetic vector potential describes},
  \href{https://doi.org/10.1119/1.11298}{\emph{Am. J. Phys} {\bfseries 46}
  (1978) 499}
  [\href{https://arxiv.org/abs/https://doi.org/10.1119/1.11298}{{\ttfamily
  https://doi.org/10.1119/1.11298}}].

\bibitem{Blake:2013owa}
M.~Blake, D.~Tong and D.~Vegh, \emph{Holographic lattices give the graviton an
  effective mass},
  \href{https://doi.org/10.1103/PhysRevLett.112.071602}{\emph{Phys. Rev. Lett.}
  {\bfseries 112} (2014) 071602}
  [\href{https://arxiv.org/abs/1310.3832}{{\ttfamily 1310.3832}}].

\bibitem{Vegh:2013sk}
D.~Vegh, \emph{Holography without translational symmetry},
  \href{https://arxiv.org/abs/1301.0537}{{\ttfamily 1301.0537}}.

\bibitem{Donos:2014cya}
A.~Donos and J.~P. Gauntlett, \emph{Thermoelectric dc conductivities from black
  hole horizons}, \href{https://doi.org/10.1007/JHEP11(2014)081}{\emph{JHEP}
  {\bfseries 11} (2014) 081} [\href{https://arxiv.org/abs/1406.4742}{{\ttfamily
  1406.4742}}].

\bibitem{Skenderis:2008dg}
K.~Skenderis and B.~C. van Rees, \emph{Real-time gauge/gravity duality:
  Prescription, renormalization and examples},
  \href{https://doi.org/10.1088/1126-6708/2009/05/085}{\emph{JHEP} {\bfseries
  05} (2009) 085} [\href{https://arxiv.org/abs/0812.2909}{{\ttfamily
  0812.2909}}].

\bibitem{Taylor:2000xw}
M.~Taylor, \emph{More on counterterms in the gravitational action and
  anomalies},  \href{https://arxiv.org/abs/hep-th/0002125}{{\ttfamily
  hep-th/0002125}}.

\bibitem{Grozdanov:2017kyl}
S.~Grozdanov and N.~Poovuttikul, \emph{Generalised global symmetries in
  holography: magnetohydrodynamic waves in a strongly interacting plasma},
  \href{https://doi.org/10.1007/JHEP04(2019)141}{\emph{JHEP} {\bfseries 04}
  (2019) 141} [\href{https://arxiv.org/abs/1707.04182}{{\ttfamily
  1707.04182}}].

\bibitem{Ammon:2017ded}
M.~Ammon, M.~Kaminski, R.~Koirala, J.~Leiber and J.~Wu, \emph{Quasinormal modes
  of charged magnetic black branes \& chiral magnetic transport}, {\emph{JHEP}
  {\bfseries 04} (2017) 067}.

\bibitem{Chesler:2009cy}
P.~M. Chesler and L.~G. Yaffe, \emph{Boost invariant flow, black hole
  formation, and far-from-equilibrium dynamics in n = 4 supersymmetric
  yang-mills theory},
  \href{https://doi.org/10.1103/PhysRevD.82.026006}{\emph{Phys. Rev.}
  {\bfseries D82} (2010) 026006}
  [\href{https://arxiv.org/abs/0906.4426}{{\ttfamily 0906.4426}}].

\bibitem{Chesler:2013lia}
P.~M. Chesler and L.~G. Yaffe, \emph{Numerical solution of gravitational
  dynamics in asymptotically anti-de sitter spacetimes},
  \href{https://doi.org/10.1007/JHEP07(2014)086}{\emph{JHEP} {\bfseries 07}
  (2014) 086} [\href{https://arxiv.org/abs/1309.1439}{{\ttfamily 1309.1439}}].

\bibitem{wilkethesis}
W.~van~der Schee, \emph{Gravitational collisions and the quark-gluon plasma},
  Ph.D. thesis, Utrecht University, 2014.

\bibitem{Janik:2017ykj}
R.~A. Janik, J.~Jankowski and H.~Soltanpanahi, \emph{Real-time dynamics and
  phase separation in a holographic first order phase transition},
  \href{https://doi.org/10.1103/PhysRevLett.119.261601}{\emph{Phys. Rev. Lett.}
  {\bfseries 119} (2017) 261601}
  [\href{https://arxiv.org/abs/1704.05387}{{\ttfamily 1704.05387}}].

\bibitem{Ecker:2018jgh}
C.~Ecker, \emph{Entanglement Entropy from Numerical Holography}, Ph.D. thesis,
  Vienna, Tech. U., 2018-09.
\newblock \href{https://arxiv.org/abs/1809.05529}{{\ttfamily 1809.05529}}.

\bibitem{boyd}
J.~Boyd, \emph{Chebyshev and Fourier Spectral Methods}. Dover Publications,
  Inc., 2000.

\bibitem{numericalrecipes}
W.~H. Press, S.~A. Teukolsky, W.~T. Vetterling and B.~P. Flannery,
  \emph{Numerical Recipes; The Art of Scientific Computing}. Cambridge
  University Press, 3\textsuperscript{rd}~ed., 2007.

\bibitem{Son:2009tf}
D.~T. Son and P.~Surowka, \emph{Hydrodynamics with triangle anomalies},
  \href{https://doi.org/10.1103/PhysRevLett.103.191601}{\emph{Phys. Rev. Lett.}
  {\bfseries 103} (2009) 191601}
  [\href{https://arxiv.org/abs/0906.5044}{{\ttfamily 0906.5044}}].

\bibitem{Bhattacharyya:2008xc}
S.~Bhattacharyya, V.~E. Hubeny, R.~Loganayagam, G.~Mandal, S.~Minwalla,
  T.~Morita et~al., \emph{Local fluid dynamical entropy from gravity},
  \href{https://doi.org/10.1088/1126-6708/2008/06/055}{\emph{JHEP} {\bfseries
  06} (2008) 055} [\href{https://arxiv.org/abs/0803.2526}{{\ttfamily
  0803.2526}}].

\bibitem{Bousso:2015mqa}
R.~Bousso and N.~Engelhardt, \emph{New area law in general relativity},
  \href{https://doi.org/10.1103/PhysRevLett.115.081301}{\emph{Phys. Rev. Lett.}
  {\bfseries 115} (2015) 081301}
  [\href{https://arxiv.org/abs/1504.07627}{{\ttfamily 1504.07627}}].

\bibitem{Sanches:2016pga}
F.~Sanches and S.~J. Weinberg, \emph{Refinement of the bousso-engelhardt area
  law}, \href{https://doi.org/10.1103/PhysRevD.94.021502}{\emph{Phys. Rev.}
  {\bfseries D94} (2016) 021502}
  [\href{https://arxiv.org/abs/1604.04919}{{\ttfamily 1604.04919}}].

\bibitem{2005JSMTE04010C}
P.~{Calabrese} and J.~{Cardy}, \emph{Evolution of entanglement entropy in
  one-dimensional systems},
  \href{https://doi.org/10.1088/1742-5468/2005/04/P04010}{\emph{JSTAT}
  {\bfseries 4} (2005) 04010}
  [\href{https://arxiv.org/abs/cond-mat/0503393}{{\ttfamily
  cond-mat/0503393}}].

\bibitem{Liu:2013iza}
H.~Liu and S.~J. Suh, \emph{Entanglement tsunami: Universal scaling in
  holographic thermalization},
  \href{https://doi.org/10.1103/PhysRevLett.112.011601}{\emph{Phys. Rev. Lett.}
  {\bfseries 112} (2014) 011601}
  [\href{https://arxiv.org/abs/1305.7244}{{\ttfamily 1305.7244}}].

\bibitem{Nishioka:2009un}
T.~Nishioka, S.~Ryu and T.~Takayanagi, \emph{Holographic entanglement entropy:
  An overview}, \href{https://doi.org/10.1088/1751-8113/42/50/504008}{\emph{J.
  Phys. A} {\bfseries 42} (2009) 504008}
  [\href{https://arxiv.org/abs/0905.0932}{{\ttfamily 0905.0932}}].

\bibitem{Hubeny:2012ry}
V.~E. Hubeny, \emph{Extremal surfaces as bulk probes in ads/cft},
  \href{https://doi.org/10.1007/JHEP07(2012)093}{\emph{JHEP} {\bfseries 07}
  (2012) 093} [\href{https://arxiv.org/abs/1203.1044}{{\ttfamily 1203.1044}}].

\bibitem{Ecker:2019ocp}
C.~Ecker, D.~Grumiller, W.~van~der Schee, M.~Sheikh-Jabbari and P.~Stanzer,
  \emph{Quantum null energy condition and its (non)saturation in 2d cfts},
  \href{https://doi.org/10.21468/SciPostPhys.6.3.036}{\emph{SciPost Phys.}
  {\bfseries 6} (2019) 036} [\href{https://arxiv.org/abs/1901.04499}{{\ttfamily
  1901.04499}}].

\bibitem{Semenoff:2011ng}
G.~W. Semenoff and K.~Zarembo, \emph{Holographic schwinger effect},
  \href{https://doi.org/10.1103/PhysRevLett.107.171601}{\emph{Phys. Rev. Lett.}
  {\bfseries 107} (2011) 171601}
  [\href{https://arxiv.org/abs/1109.2920}{{\ttfamily 1109.2920}}].

\bibitem{Gorsky:2001up}
A.~S. Gorsky, K.~A. Saraikin and K.~G. Selivanov, \emph{Schwinger type
  processes via branes and their gravity duals},
  \href{https://doi.org/10.1016/S0550-3213(02)00095-0}{\emph{Nucl. Phys.}
  {\bfseries B628} (2002) 270}
  [\href{https://arxiv.org/abs/hep-th/0110178}{{\ttfamily hep-th/0110178}}].

\bibitem{Ambjorn:2011wz}
J.~Ambjorn and Y.~Makeenko, \emph{Remarks on holographic wilson loops and the
  schwinger effect},
  \href{https://doi.org/10.1103/PhysRevD.85.061901}{\emph{Phys. Rev.}
  {\bfseries D85} (2012) 061901}
  [\href{https://arxiv.org/abs/1112.5606}{{\ttfamily 1112.5606}}].

\bibitem{Kawai:2013xya}
D.~Kawai, Y.~Sato and K.~Yoshida, \emph{Schwinger pair production rate in
  confining theories via holography},
  \href{https://doi.org/10.1103/PhysRevD.89.101901}{\emph{Phys. Rev.}
  {\bfseries D89} (2014) 101901}
  [\href{https://arxiv.org/abs/1312.4341}{{\ttfamily 1312.4341}}].

\bibitem{Yee:2009vw}
H.-U. Yee, \emph{Holographic chiral magnetic conductivity},
  \href{https://doi.org/10.1088/1126-6708/2009/11/085}{\emph{JHEP} {\bfseries
  11} (2009) 085} [\href{https://arxiv.org/abs/0908.4189}{{\ttfamily
  0908.4189}}].

\bibitem{Sato:2013pxa}
Y.~Sato and K.~Yoshida, \emph{Holographic description of the schwinger effect
  in electric and magnetic fields},
  \href{https://doi.org/10.1007/JHEP04(2013)111}{\emph{JHEP} {\bfseries 04}
  (2013) 111} [\href{https://arxiv.org/abs/1303.0112}{{\ttfamily 1303.0112}}].

\bibitem{Ammon:2016fru}
M.~Ammon, S.~Grieninger, A.~Jimenez-Alba, R.~P. Macedo and L.~Melgar,
  \emph{Holographic quenches and anomalous transport},
  \href{https://doi.org/10.1007/JHEP09(2016)131}{\emph{JHEP} {\bfseries 09}
  (2016) 131} [\href{https://arxiv.org/abs/1607.06817}{{\ttfamily
  1607.06817}}].

\bibitem{Haack:2018ztx}
M.~Haack, D.~Sarkar and A.~Yarom, \emph{Probing anomalous driving},
  \href{https://doi.org/10.1007/JHEP04(2019)034}{\emph{JHEP} {\bfseries 04}
  (2019) 034} [\href{https://arxiv.org/abs/1812.08210}{{\ttfamily
  1812.08210}}].

\bibitem{Bali:2011qj}
G.~S. Bali, F.~Bruckmann, G.~Endrodi, Z.~Fodor, S.~D. Katz, S.~Krieg et~al.,
  \emph{The qcd phase diagram for external magnetic fields},
  \href{https://doi.org/10.1007/JHEP02(2012)044}{\emph{JHEP} {\bfseries 02}
  (2012) 044} [\href{https://arxiv.org/abs/1111.4956}{{\ttfamily 1111.4956}}].

\bibitem{Zhong:2014cda}
Y.~Zhong, C.-B. Yang, X.~Cai and S.-Q. Feng, \emph{A systematic study of
  magnetic field in relativistic heavy-ion collisions in the rhic and lhc
  energy regions}, \href{https://doi.org/10.1155/2014/193039}{\emph{Adv. High
  Energy Phys.} {\bfseries 2014} (2014) 193039}
  [\href{https://arxiv.org/abs/1408.5694}{{\ttfamily 1408.5694}}].

\bibitem{Pang:2016yuh}
L.-G. Pang, G.~Endrődi and H.~Petersen, \emph{Magnetic-field-induced squeezing
  effect at energies available at the bnl relativistic heavy ion collider and
  at the cern large hadron collider},
  \href{https://doi.org/10.1103/PhysRevC.93.044919}{\emph{Phys. Rev.}
  {\bfseries C93} (2016) 044919}
  [\href{https://arxiv.org/abs/1602.06176}{{\ttfamily 1602.06176}}].

\bibitem{Gursoy:2018yai}
U.~Gürsoy, D.~Kharzeev, E.~Marcus, K.~Rajagopal and C.~Shen,
  \emph{Charge-dependent flow induced by magnetic and electric fields in heavy
  ion collisions},  \href{https://arxiv.org/abs/1806.05288}{{\ttfamily
  1806.05288}}.

\bibitem{Ye:2018jwq}
Y.~J. Ye, Y.~G. Ma, A.~H. Tang and G.~Wang, \emph{Effect of magnetic fields on
  pairs of oppositely charged particles in ultrarelativistic heavy-ion
  collisions}, \href{https://doi.org/10.1103/PhysRevC.99.044901}{\emph{Phys.
  Rev. C} {\bfseries 99} (2019) 044901}.

\bibitem{Jimenez-Alba:2014iia}
A.~Jimenez-Alba, K.~Landsteiner and L.~Melgar, \emph{Anomalous magnetoresponse
  and the stückelberg axion in holography},
  \href{https://doi.org/10.1103/PhysRevD.90.126004}{\emph{Phys. Rev.}
  {\bfseries D90} (2014) 126004}
  [\href{https://arxiv.org/abs/1407.8162}{{\ttfamily 1407.8162}}].

\bibitem{Klebanov:2002gr}
I.~R. Klebanov, P.~Ouyang and E.~Witten, \emph{A gravity dual of the chiral
  anomaly}, \href{https://doi.org/10.1103/PhysRevD.65.105007}{\emph{Phys. Rev.}
  {\bfseries D65} (2002) 105007}
  [\href{https://arxiv.org/abs/hep-th/0202056}{{\ttfamily hep-th/0202056}}].

\bibitem{Grozdanov:2016tdf}
S.~Grozdanov, D.~M. Hofman and N.~Iqbal, \emph{Generalized global symmetries
  and dissipative magnetohydrodynamics},
  \href{https://doi.org/10.1103/PhysRevD.95.096003}{\emph{Phys. Rev.}
  {\bfseries D95} (2017) 096003}
  [\href{https://arxiv.org/abs/1610.07392}{{\ttfamily 1610.07392}}].

\bibitem{Grozdanov:2018fic}
S.~Grozdanov, A.~Lucas and N.~Poovuttikul, \emph{Holography and hydrodynamics
  with weakly broken symmetries},
  \href{https://doi.org/10.1103/PhysRevD.99.086012}{\emph{Phys. Rev.}
  {\bfseries D99} (2019) 086012}
  [\href{https://arxiv.org/abs/1810.10016}{{\ttfamily 1810.10016}}].

\bibitem{Armas:2018atq}
J.~Armas and A.~Jain, \emph{Magnetohydrodynamics as superfluidity},
  \href{https://doi.org/10.1103/PhysRevLett.122.141603}{\emph{Phys. Rev. Lett.}
  {\bfseries 122} (2019) 141603}
  [\href{https://arxiv.org/abs/1808.01939}{{\ttfamily 1808.01939}}].

\bibitem{Armas:2018zbe}
J.~Armas and A.~Jain, \emph{One-form superfluids \& magnetohydrodynamics},
  \href{https://doi.org/10.1007/JHEP01(2020)041}{\emph{JHEP} {\bfseries 01}
  (2020) 041} [\href{https://arxiv.org/abs/1811.04913}{{\ttfamily
  1811.04913}}].

\end{thebibliography}\endgroup

\end{document}